\documentclass[a4paper,11pt]{article}
\usepackage{jcappub}

\usepackage{lineno}

\usepackage[utf8]{inputenc} 

\usepackage{calligra}
\usepackage{graphicx,mathtools,amssymb,amsmath,amsthm,amsfonts,epsfig,epsf}
\usepackage[outdir=./]{epstopdf}
\usepackage[dvipsnames]{xcolor}

\hypersetup{pdfstartview=}	
\usepackage{tensor}	
\usepackage{csquotes}
\usepackage{mathrsfs}
\usepackage{float}
\usepackage[normalem]{ulem}
\usepackage{multirow}
\usepackage{subfig}
\usepackage{ dsfont }
\usepackage[normalem]{ulem}

\usepackage[font=small,labelfont=bf, justification=Justified,format=plain]{caption}

\title{Deep learning inference of the neutron star equation of state}

\author[a]{Giulia Ventagli}
\author[a]{and Ippocratis D. Saltas}
\affiliation[a]{CEICO, Institute of Physics of the Czech Academy of Sciences, Na Slovance 2, 182 21 Praha 8, Czechia}

\emailAdd{ventagli@fzu.cz}
\emailAdd{saltas@fzu.cz}

\abstract{
We present a pipeline to infer the equation of state of neutron stars from observations based on deep neural networks. In particular, using the standard (deterministic), as well as Bayesian (probabilistic) deep networks,  we explore how one can infer the interior speed of sound of the star given a set of mock observations of total stellar mass, stellar radius and tidal deformability. We discuss in detail the construction of our simulated dataset of stellar observables starting from the solution of the gravitational equations, as well as the relevant architectures for the deep networks, along with their performance and accuracy. We further explain how our pipeline is capable to detect a possible QCD phase transition in the stellar core. Our results show that deep networks offer a promising tool towards solving the inverse problem of neutron stars, and the accurate inference of their interior from future stellar observations.}

\begin{document}
\maketitle
\flushbottom

\section{Introduction}

Neutron stars provide a remarkable tool to probe the fundamental nature of cold, dense, strongly-interacting matter~\cite{Baym:2017whm}, thanks to the densities of their core which can reach values not accessible in terrestrial laboratories. The surge of neutron star observations in the last decade from ray and radio emissions, as well as from gravitational waves~\cite{Ozel:2010fw,Steiner:2010fz,Steiner:2012xt,Lattimer:2013hma,Steiner:2014pda,Ozel:2015fia,Steiner:2015aea,Ozel:2016oaf,Miller:2016pom,Alvarez-Castillo:2016oln, Steiner:2017vmg,Ascenzi:2024wws,Antoniadis:2013pzd,NANOGrav:2017wvv,Fonseca:2021wxt,Riley:2021pdl,Miller:2021qha,LIGOScientific:2018hze,LIGOScientific:2020aai} has provided astrophysical inferences on neutron star masses, radii and tidal deformabilities, allowing us to further constraint the equation of state (EOS). Hence, achieving an optimal way to infer EOS' properties from observation is a crucial task. A way to tackle such issue is to perform a statistical analysis in the framework of Bayesian inference, as was developed in Refs.~\cite{Steiner:2010fz,Steiner:2012xt,Ozel:2015fia}. 

Recently, machine learning techniques have also been used to predict EOS' features from observations~\cite{Fujimoto:2017cdo,Fujimoto:2019hxv,Ferreira:2019bny,Morawski:2020izm,Traversi:2020dho,Fujimoto:2021zas,Krastev:2021reh,Soma:2022qnv,Han:2022sxt,Soma:2022vbb,Soma:2023fyi,Thete:2022drz,Farrell:2022lfd,Ferreira:2022nwh,Goncalves:2022smd,Anil:2020lch,Chatterjee:2023ecc,Krastev:2023fnh,Farrell:2023ojk,Zhou:2023cfs,Carvalho:2023ele,Carvalho:2024kgf,Ferreira:2024rnf,Fujimoto:2024cyv,Brandes:2024vhw,Thakur:2024mxs}. Roughly, two different approaches have been followed in the literature: reconstructing the EOS from either stellar masses, stellar radii and tidal deformabilities~\cite{Fujimoto:2017cdo,Fujimoto:2019hxv,Morawski:2020izm,Traversi:2020dho,Fujimoto:2021zas,Soma:2022qnv,Han:2022sxt,Soma:2022vbb,Soma:2023fyi,Farrell:2022lfd,Goncalves:2022smd,Chatterjee:2023ecc,Farrell:2023ojk,Zhou:2023cfs,Carvalho:2023ele,Ferreira:2024rnf,Fujimoto:2024cyv,Brandes:2024vhw,Thakur:2024mxs}, or focusing directly on the nuclear matter saturation properties~\cite{Ferreira:2019bny,Krastev:2021reh,Thete:2022drz,Ferreira:2022nwh,Anil:2020lch,Krastev:2023fnh,Carvalho:2023ele,Carvalho:2024kgf}. Our work falls into the first category. In particular, we construct a pipeline for supervised deep neural networks which allow us to predict the EOS from mock observational data composed of masses, radii and tidal deformabilities (or equivalently, tidal Love number). We employ both conventional (deterministic) and Bayesian (probabilistic) architectures. The latter utilises variational layers, where the weights and biases are drawn from a learnable posterior distribution, whereas in the deterministic network the weights and biases are point estimates. To the best of our knowledge, the employment of variational layers to account for epistemic uncertainty in the dataset is a novelty. While Bayesian neural networks have previously been used in the literature, our approach is inherently different. In particular, in Refs.~\cite{Carvalho:2023ele,Carvalho:2024kgf}, the authors employed a Bayesian approach to minimise a log-likelihood related to neutron star observables, using a fundamentally different EOS parametrization to ours. What is more, Ref.~\cite{Brandes:2024vhw} studied inferences of the EOS directly from telescope spectra employing the method of normalising flow for the reconstruction of the  probability density functional.

We work in the context of a new prescriptions for modelling EOSs, following Ref.~\cite{Ventagli:2024cho}. We assume that neutron stars are described by an EOS composed of three regions. At low-density we employ a realistic tabulated EOS (in this work we choose either AP4 or SLy), while at high-density we introduce a QCD phase by means of a speed-of-sound parametrization~\cite{Tews:2018iwm}. Most importantly, we also allow for a vacuum energy phase transition in the stellar core. The latter is a key novelty of our work. A vacuum energy shift, which can be interpreted as a change in the underlying value of the cosmological constant, is to be expected in the core of the star, where the pressure can be sufficiently high to trigger the QCD phase transition. This mechanism has been proposed in Ref.~\cite{Bellazzini:2015wva,Csaki:2018fls,Ventagli:2024cho} to study neutron stars as novel probes for the cosmological constant problem. While it is challenging to test the evolution of the vacuum energy in the Universe, one can focus on scenarios where the structure of the vacuum is rearranged with sizable shift in the vacuum energy, as in the core of neutron stars. Hence, an imprint in the stellar properties can shed some light into the gravitational effect of such transition.

The aim of our work can be summarised as follows:
\begin{itemize}
    \item The construction of a training and testing dataset comprised of stellar masses, stellar radii and tidal deformabilities, or equivalently tidal Love numbers, from a set of EOSs.
    \item The determination of the low-density EOS with a  classification deep network.
    \item The prediction of the vacuum energy shift and the speed-of-sound profile through a regression analysis with both a deterministic and a Bayesian neural network.
\end{itemize}

The paper is structured as follows: in Section~\ref{Sec:NS} we describe the modelling of the neutron star interior and the numerical pipeline we use to solve the equations which leads to our training and testing datasets. In Section~\ref{Sec:NN} we describe our deep-learning pipeline, which we split into an analysis based on a deterministic and a Bayesian deep network respectively. In Section~\ref{Sec:RealData}, we apply our network to real observational data. We summarise in Section \ref{Sec:Conclusions}. The numerical data and codes of our analysis are {\bf available online} at Ref.~\cite{ventagli_2024_11216586}.

\begin{figure}[h]
\centering
\includegraphics[scale=0.5]{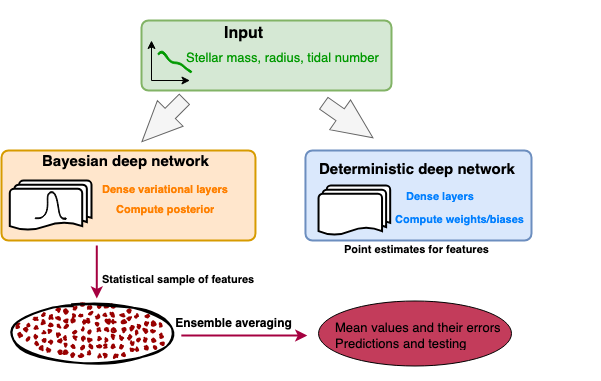}
\caption{A sketch showing an overview of the deep networks we use, i.e. a ``standard" deterministic and a probabilistic network respectively. The input data to our network are simulated values for the mass, radius and tidal number of the given neutron star, computed according to the procedure explained in the paper, whereas the output is predominantly a prediction of the speed of sound. Our probabilistic deep network accounts for the epistemic uncertainty in the dataset, by using learnable probability functions for the weights and biases of the network, as explained in Sec. \ref{Sec:Bayesian}. }
\end{figure}

\section{Neutron star modelling}\label{Sec:NS}

In this work, we consider static, spherically symmetric stars described by the metric ansatz
\begin{equation}\label{eq:eqconfig}
    ds^2 = g^{(0)}_{\mu\nu} dx^\mu dx^\nu = -e^{\nu(r)} dt^2 + e^{\mu(r)} dr^2 + r^2 d\Omega^2\ ,
\end{equation}
and we assume that matter is described by a perfect fluid with a stress-energy tensor given by $T_{\mu\nu}^{(0)} = (\epsilon + p) u_\mu u_\nu + p\,g^{(0)}_{\mu\nu}$, where $u_\mu=(-e^{\nu(r)/2},0,0,0)$ is the fluid's four velocity, $p$ is the pressure and $\epsilon$ is the energy density. The pressure is directly related to the energy density by a barotropic EOS, $\epsilon=\epsilon(p)$. One determines the metric functions $\nu(r)$ and $\mu(r)$, and the pressure $p$ as solutions of the Tolman-Oppenheimer-Volkoff (TOV) system of equations~\cite{Tolman:1939jz,Oppenheimer:1939ne}
\begin{equation}\label{eq:tov1}
\begin{split}
    m'(r) &= 4 \pi r^2 \epsilon(r)\ ,\\   
    p'(r) &= -\frac{p(r)+\epsilon(r)}{r[r-2 m(r)]} G [m(r)+4\pi r^3 p(r)],\\
    \nu'(r) &= -\frac{2p'(r)}{p(r)+\epsilon(r)}  \ ,
\end{split}
\end{equation}
where primes denote differentiation with respect to the radial 
coordinate $r$. One solves the TOV equations by imposing appropriate boundary condition at the centre of the star ($r=r_0$), and integrating outward to the surface of the star ($r=R$), where for the pressure and mass we have $p(R)=0$ and $m(R)=M$, with $M$ and $R$ the total mass and radius respectively.

We assume that the neutron star is immersed in an external static and 
quadrupolar tidal field, $\mathcal{E}_{ij}$. At leading order, the quadrupole moment, $Q_{ij}$, induced on the star is linearly proportional to $\mathcal{E}_{ij}$. The coupling between such quantities is given by the 
dimensionful tidal deformability $\lambda$, that is
\begin{equation}
    Q_{ij} = -\lambda \mathcal{E}_{ij}=-\frac{2}{3}k_2R^5\mathcal{E}_{ij}\ ,\label{eq:adiabaticapprox}
\end{equation}
where in the last step we have introduced the dimensionless $l=2$ Love number, $k_2$~\cite{Thorne:1997kt,Hinderer:2007mb,Hinderer:2009ca,Binnington:2009bb,Damour:2009vw}. The latter can be computed by considering a linear perturbation to the background solution~\eqref{eq:eqconfig},
\begin{equation}
    g_{\mu\nu} = g^{(0)}_{\mu\nu} + h_{\mu\nu}(r,\theta,\phi) \ ,
\end{equation}
Under an expansion in spherical harmonics, $Y_{lm}(\theta,\phi)$, the ($l=2$) metric perturbations and the fluid energy-momentum read as 
\begin{equation}
 h_{\mu\nu} = e^{\nu(r)} H(r) Y_{20}(\theta,\phi) dt^2 + e^{\mu(r)} H(r) Y_{20}(\theta,\phi) dr^2 + r^2 K(r) Y_{20}(\theta,\phi) d\Omega^2\ ,\label{eq:metrich}
\end{equation}
\begin{equation}
T_{\mu\nu}=T_{\mu\nu}^{(0)}+\delta T_{\mu\nu},
\end{equation}
with 
$
\delta T^0_0=-\delta \epsilon(r) Y_{20}(\theta,\phi), \,  
\delta T^i_i=\delta p(r) Y_{20}(\theta,\phi)\ .
$
Employing Einstein field's equations, we can express $K(r)$ 
as a function of $H(r)$, $K'(r) = H' + H(r) \nu'(r)$, allowing us to cast the perturbations in terms of a single second order differential equation
\begin{multline}\label{eq:diffH}
    H''(r) -\frac{2}{r} e^\mu \big[ 2\pi G r^2 (\epsilon-p)-1\big]H'(r) 
    -2 e^\mu \bigg\{ \frac{3}{r^2} -2\pi G [5\epsilon +9 p + (\epsilon+p) f ] \\
    +2 G^2 e^\mu \left( \frac{m(r)}{r^2}+4\pi r p \right)^2 \bigg\} H(r) = 0\ ,
\end{multline}
where $f=d\epsilon/dp$. Introducing $\beta(r) \equiv dH/dr$, we solve Eq.~\eqref{eq:diffH} integrating from the stellar center up to the surface with boundary conditions. In the inner boundary we impose 
\begin{equation}
 H(r_0) \sim a_0 r_0^2 \quad\ ,\quad
\beta(r_0) = 2 a_0 r_0\ ,
\end{equation}
with $a_0$ a numerical constant~\footnote{The value of this constant is irrelevant, as it cancels out in the final calculation of $k_2$.}, while at the surface $H(r)$ is matched to the vacuum solution with 
$T_{\mu\nu}=0$. Outside the star, one finds that $H(r)$ can be expressed analytically in terms of the associated Legendre functions ${\cal Q}_{22}$ and ${\cal P}_{22}$~\cite{Thorne:1967,Hinderer:2007mb,Olver:2010}
\begin{equation}
H(r\ge R)=c_{1} {\cal P}_{22}\left(r/M-1\right)+
c_{2} {\cal Q}_{22}\left(r/M-1\right)\ ,\label{eq:HrR}
\end{equation}
where $c_{1,2}$ are two constants of integration computed through the matching. The $g_{tt}$ metric component at spatial infinity can be written as the sum of a decaying ($Q^2_2(r/m-1)\sim r^{-3}$ ) and a growing ($P^2_2(r/m-1)\sim r^2$) mode, which is matched with the corresponding asymptotic form written in the so-called ACMC coordinates~\cite{Thorne:1980ru}
\begin{equation}\label{eq:metrExp}
    -\frac{1+g_{tt}}{2} = -\frac{m}{r}-\frac{3}{2}\frac{Q_{ij}}{r^3}n^in^j+...+\frac{\mathcal{E}_{ij}}{2}r^2 n^i n^j+...,    
\end{equation}
where $n^i=x^i/r$ \cite{Thorne:1980ru}. This procedure 
allows to identify the growing (tidal field) and decaying (quadrupole deformation) solutions which can be related to the two integration constants. Finally, we can express the Love number as 
\begin{multline}
    k_2 = \frac{8 C^5}{5}(1 - 2C)^2(2 + 2C(y-1) -1)
    \bigg\{ 2 C \big[6 - 3y + 3C(5y - 8)\big] \\
    + 4C^3 \big[13 - 11 y + C(3y - 2) + 2C^2(1 + y)\big] \\
    + 3 (1 - 2C)^2 \big[ 2 - y + 2C(y - 1) \big] \text{ln}(1 - 2C) \bigg\}^{-1},
\end{multline}
where $C \equiv M/R$ is the stellar compactness, and $y \equiv R H'(R)/H(R)$, with derivatives understood to be on the stellar surface.

In summary, given a choice for the EOS, solving the system in Eq.~\eqref{eq:tov1} together with Eq.~\eqref{eq:diffH} allows us to extract the total stellar radius $R$, mass $M$ and dimensionless $l=2$ Love number $k_2$ respectively. Schematically,
\begin{equation}\label{eq:tov}
(\text{EOS}) \mapsto (M,\,R,\,k_2).
\end{equation}

It is evident that the specific EOS chosen plays a crucial role. In this work, we adopt the formalism used in Ref.~\cite{Ventagli:2024cho}. We assume that the neutron star structure is described by an EOS composed of three regions:
\begin{description}
\item[Region I] A low-density part up to a threshold density $\rho_\text{tr}=2\rho_0$, where $\rho_0\simeq 2.7 \times 10^{14}\, \text{g}/\text{cm}^{3}$ is the saturation density. We adopt nucleonic (tabulated) EOSs. Specifically, we choose two models commonly used in the literature, namely, the SLy EOS~\cite{Douchin:2001sv} and the AP4 EOS~\cite{Akmal:1998cf}. Even though such models are obtained by different methodologies and calculation schemes, they both describe soft nuclear matter and are consistent with current astrophysical observations~\cite{Miller:2021qha,Miller:2019cac,Riley:2019yda,Riley:2021pdl,LIGOScientific:2020aai,LIGOScientific:2018hze}. 
\item[Region II] A high-density regime for $\rho>\rho_\text{tr}$, where the nuclear matter experiences a QCD phase transition. We employ a phenomenological EOS using an agnostic speed-of-sound model~\cite{Tews:2018kmu,Tews:2018iwm}. In particular, we construct a linear approximation of the speed of sound $c_s^2 \equiv \frac{\partial p(\epsilon)}{\partial \epsilon}$, randomly sampling seven reference points $(\rho,c_s)$ within $\rho\in(\rho_\text{tr}, 12\,\rho_0)$ and $c_s\in(0, 1)$, and connecting them by linear segments. Note that for the seventh point, we always assume the mass density to be fixed, i.e. $\rho_7=12\,\rho_0$, while we allow the speed of sound to acquire random values. Even though such large value for $\rho_7$ is unlikely to be reached by neutron stars on the stable branch, we chose to include it in our dataset to be in line with the initial assumptions of Ref.~\cite{Tews:2018kmu}. We assume causality is preserved, thus we discard models with $c_s > 1$ to avoid unphysical EOSs.
\item[Region III] a vacuum energy phase transition in the inner stellar core, for values of the pressure larger than $p_c=(200\, \text{MeV})^4$. Indeed, deep within the stellar core, the local pressure can be sufficiently high to trigger the QCD phase transition. This would generate a shift in the underlying value of the vacuum energy. One can interpret this shift in energy as a new effective cosmological constant term $\Lambda$ contributing to the pressure, energy density and mass density of the the new exotic phase. Prior to the vacuum energy transition, the gluon condensate is thought to contribute negatively to the overall vacuum energy by an amount $-0.0034~(\text{GeV})^4$~\cite{Donoghue:2017vvl}. This is the standard value usually quoted in the literature, despite some mild controversy~\cite{Holdom:2007gg,Holdom:2009ma}. Regardless of the precise details, it is reasonable to expect the vacuum energy to {\it increase} by $(\mathcal{O}(100) \text{MeV})^4$ as we transition towards the inner core of the neutron star~\cite{Ventagli:2024cho}.
Nevertheless, we allow for both positive and negative values of $\Lambda$, to consider the possibility of an unknown particle physics mechanism screening some or all of the underlying change in vacuum energy. We follow Ref.~\cite{Ventagli:2024cho}, and we choose 10 values of $\Lambda$ in the range $\big(-(194\, \text{MeV})^4, (194\, \text{MeV})^4\big)$, including the case where no phase transition develops in the core, i.e. $\Lambda = 0$.
\end{description}

\section{Deep learning pipeline}\label{Sec:NN}

In this section, we describe the concrete setup of our deep learning network. Our scope is to constrain the neutron star EOS from observable properties, such as the mass, the radius, and the tidal deformability of the star, or equivalently its $l=2$ dimensionless tidal Love number. Doing so corresponds to solving the ``inverse'' problem we delineated in Sec.~\ref{Sec:NS} and described by Eq.~\eqref{eq:tov}, that is, we aim at computing 
\begin{equation}\label{eq:inverse}
    (M,\,R,\,k_2) \mapsto (\text{EOS}).
\end{equation}
While it is known that such inverted mapping exists~\cite{Lindblom:2014sha}, one cannot practically retrieve it. First of all, we cannot form a continuous curve from the $M$-$R$ observational data or the $k_2$-$M$ ones. Furthermore, observational data have uncertainties and are described by distributions, hence one is not able to reconstruct the genuine $M$-$R$ and $k_2$-$M$ relations (see Ref.~\cite{Fujimoto:2021zas} for a further discussion on the issue). Lastly, some initial information about the EOS at low density should be known beforehand. We tackle this challenge with an alternative approach, similar to Ref.~\cite{Fujimoto:2021zas}. In particular, instead of seeking the unique inverse mapping, i.e. Eq.~\eqref{eq:inverse}, we infer the most likely EOS from observations data including their uncertainties through  a regression analysis with deep-learning techniques. We do this at two steps. We first consider a ``standard" (deterministic) deep network. In this case, the weights and biases of the network at training are point estimates, which implies that once the network is trained, a given input will always output the same prediction, modulo the inherit stochasticity in deep networks.
Our second approach will be a Bayesian deep network, able to capture the epistemic uncertainty in the data. In this approach, the weights and biases at each node are drawn from learnable probability density functions, and the output of the network will be a statistical realisation of the given input. 
Before we dive into the description of the deep architecture and their results, we first discuss the construction of the training and validation dataset.

\subsection{Training and validation dataset}\label{Sec:dataset}

In order to train and test both the deterministic and the Bayesian neural networks, we first generate mock observational data as input data. We first solve the TOV equations~\eqref{eq:tov1} for randomly generated EOSs (see Sec.~\ref{Sec:NS}), obtaining for each EOS a $M$-$R$ and a $k_2$-$M$ curve respectively. Note that, in order to generate data in agreement with the heaviest neutron star observed so far (PSR J0952-0607~\cite{Romani:2022jhd}), which has a mass $M_\text{max} = (2.35\pm 0.17)M_\odot$, we discard all EOSs which predict a maximum mass not included in this interval. For each EOS, we then randomly sample $N$ mock observation points $(M_i,R_i,k_{2,i})$, assuming a uniform distribution of neutron star masses in the interval $(M_\text{min}, M_\text{max})$, where $M_\text{min}$ is the minimum mass produced by the given EOS~\footnote{The choice of this interval allow us to discard unstable branches of the $M$-$R$ curve, while associating each $M$ to a single value of $R$ and $k_2$.}. We initially chose $N=15$, following Ref.~\cite{Fujimoto:2021zas}, but we also took into consideration the case of a larger number of mock data for a given EOS, i.e. we consider $N=20$ and $N=30$. As we will discuss in Sec.~\ref{Sec:deterministic}, allowing for a larger set of observation points for each $M$-$R$ and $k_2$-$M$ curve increases the accuracy of our network. 

To simulate observational uncertainties, we inject Gaussian noise into our mock data, i.e
\begin{align}\label{eq:injnoise}
O_i \to O_i + \mathcal{N}(\sigma_{O_i}),
\end{align}
with $O = \{M_i, R_i, k_{2_i} \}$.
We choose $\sigma_M=0.1\, M_\odot$, $\sigma_R=0.5\, \text{km}$, and $\sigma_{k_2}=0.05$ respectively. One could also allow for the standard deviations to vary for each point, further allowing the mock data to better mimic true observations. In this work, we did not consider this scenario, leaving this for future study. To obtain a sufficiently large training and validation dataset, we repeat $n$ times the injection of noise for each EOS, generating in this way batches of noised data. For the results presented in this work, we chose $n=100,\, 200,\,300$. For larger numbers of repetition, the accuracy of our neural networks increases.

As we discussed in Sec.~\ref{Sec:NS}, we generate $2 \times 10^4$ EOSs, of which only $1491$ survive the maximum mass test. For each set of points $N \times (M^\ast_i,R^\ast_i,k^\ast_{2,i})$,  we create $n$ different realisations through noise injection, leading to $N \times n \times 3$ points in total for each EOS. 
All values are normalized appropriately by employing $M^\ast_i/M_\text{norm}$, and $R^\ast_i/R_\text{norm}$ with $M_\text{norm}=3\,M_\odot$ and $R_\text{norm}=20\,\text{km}$ for masses and radii respectively, while for the tidal Love number we find the minimum and maximum value for the whole dataset, $k^\ast_{2,\text{min}}$ and $k^\ast_{2,\text{max}}$, and we normalize each $k^\ast_{2,i}$ by using $(k^\ast_{2,i}+|k^\ast_{2,\text{min}}|)/(k^\ast_{2,\text{max}}+|k^\ast_{2,\text{min}}|)$. 
The renormalization process is an important step to render the learning process faster.

As output data, for each set of points $N \times (M^\ast_i,R^\ast_i,k^\ast_{2,i})$, we store the EOS information: the corresponding low density EOS (AP4 or SLy), the value of the vacuum energy phase transition, the 7 values of mass density and speed of sound for the agnostic speed-of-sound model. We appropriately renormalize all values to ensure they belong in the (0,1) interval.

Finally, we shuffle the dataset before training to reduce overfitting, and we also split it into a $80 \%$ training and $20 \%$ validation part respectively.

\subsection{Deterministic deep network}\label{Sec:deterministic}

Here we construct a ``standard" (deterministic) deep network, where the weights and biases of the model are computed as point estimates during the training procedure. Our model needs to be able to predict $7 \times (\rho, c_s)$ points to describe the speed-of-sound parametrization and the value of the vacuum energy shift. Hence, the output data of the neural network consist of 15 parameters; 7 values of mass density, 7 values of speed of sound, and 1 value for the vacuum energy shift. However, as we discussed in Sec.~\ref{Sec:NS}, the EOSs we considered are generated with two different choices of low density EOS, namely AP4 and SLy, thus our neural network should also distinguish between these two classes. To achieve this goal we create two distinct models, a classification network which predicts the low-density EOS, and a regression model which predicts the remaining 15 parameters. 

Our input is the data described in Sec.~\ref{Sec:dataset}. We first test both the classification and the regression networks for input data composed of $M$-$R$ data only. We do so in order to test our approach on a simpler problem and compare our results against those of Ref.~\cite{Fujimoto:2021zas}. We then tackle the full problem with a neural network that takes as input $M$-$R$-$k_2$ data.

Let us first focus on the classification model, which task is to associate the input data with a low-density EOS. We tested several neural networks with different input data and setups. The input data can vary based on whether we consider a restricted $M$-$R$ dataset, a full $M$-$R$-$k_2$ dataset, the number ($n$) of noise injections we perform on each of the latter doublet/triplets, and the size of the observation set ($N$). 

For our study we used the Python library {\it Keras} with {\it TensorFlow} as a backend. We considered models with a diverse number of hidden layers, starting from a single hidden layer to at most 3, as well as different number of nodes for each layer. For every hidden layer we choose the {\tt ReLU} activation function, i.e. $\sigma(x)=\text{max}\{0,x\}$, and we initialize the kernel with the {\tt He normal} initialization, which draws samples from a truncated normal distribution centered in 0 with a standard deviation defined as $\sigma = \sqrt{2/\text{fan}\_\text{in}}$ where $\text{fan}\_\text{in}$ is the number of input units in the weight tensor~\cite{He:2015dtg}. The output layer has $2$ nodes, corresponding to the two EOS classes (AP4 and SLy), and we implement the {\tt sigmoid} activation function, namely $\sigma(x) = 1/(1+e^{-x})$, commonly used for classification models. We also implement the {\tt Glorot uniform} distribution as initializer, which draws samples from a uniform distribution within $(-\sqrt{6/(\text{fan}\_\text{in}+\text{fan}\_\text{out})},\sqrt{6/(\text{fan}\_\text{in}+\text{fan}\_\text{out})})$, where $\text{fan}\_\text{out}$ is the number of output units. We used the {\tt Adam} fitting algorithm~\cite{Kingma:2014vow}, together with the {\tt binary\_crossentropy} loss function. The latter is defined as 
\begin{equation}
    l_{\tt bin\_crossentropy}= -\frac{1}{N}  \sum_{i=1}^{N} [ y_i \cdot \text{log}(p_i)  + (1-y_i) \cdot
 \text{log}(1-p_i)],
\end{equation}
where $y$ is the actual class label, and $p_i$ is the predicted probability of the data belonging to the class. We also choose {\tt binary\_accuracy} as metric. We trained each model for $100$ epochs, and we found that a batch size of $100$ is optimal for most configurations. However, for most of the cases where $n=100$, a batch size of $50$ was preferred. We summarize the most efficient models we trained in tables I and II in the supplementary material available at Ref.~\cite{ventagli_2024_11216586}, for the $M$-$R$ input model and the model with $M$-$R$-$k_2$ respectively. Since we do not fix the random-number seed of the network, each time the same model is trained from the start it exhibits slightly different loss and accuracy at the end of training. Hence, we trained each network $20$ times, and report the mean binary accuracy and validation binary accuracy with their respective standard deviation. We can see that the model accuracy increases with the number of hidden layers, the number of nodes, the number of noise injections ($n$) and the size of the observation set ($N$). However, for high values of number of noise injections and large observations set, training networks with more than a single hidden layers and a large number of nodes is computationally demanding. For the $M$-$R$ data, the best accuracy we can reach is $0.889\pm0.001$ with 2 hidden layers of 215 and 200 nodes respectively, for a dataset with $n=300$ and $N=30$; while for the $M$-$R$-$k_2$ the best accuracy we obtain is $0.878\pm0.003$ for a single hidden layer network with 270 nodes, with the dataset corresponding to $n=300$ and $N=30$. We stress that, while a classification model is useful to distinguish between the two different low-density EOSs from which we built our dataset, this does not allow for a detailed analysis on the EOS properties in such regime, where uncertainties can still be large. We intend to improve this analysis in a future work by either allowing for a wider range of low-density EOSs from those commonly used in the literature, or by extending the analysis of the regression model to take into account this region as well.

Let us now focus on the regression model, which aims to determine the remaining 15 parameters of the EOS. As in the previous case, we tested different neural networks varying the same parameters as above. 
We developed our models using the Python library {\it Keras} with {\it TensorFlow}. For every network we tested, we only allowed for a single hidden layer, while varying its number of nodes. We have seen that including more layers does not improve the efficiency of the model, while it requires more computational time. We also chose the {\tt ReLU} activation function, with the {\tt He normal} distribution as kernel initializer. The output layer consists of 15 nodes, corresponding to the 15 EOS parameters (7 $\rho$, 7 $c_s$, and $\Lambda$), together with the {\tt sigmoid} activation function. The latter ensures that the speed of sound is bounded in the causal range $c_s \in (0,1)$~\footnote{Note that, as a result of the renormalization process discussed in Sec.~\ref{Sec:dataset}, all parameters, not only the speed of sound, are bounded in the interval $(0,1)$.}. We use the {\tt Adam} fitting algorithm~\cite{Kingma:2014vow}, and we employed the {\tt msle} loss function, that is
\begin{equation}
    l_{\tt msle}(y,\hat{y}) = \frac{1}{N} \sum_{i=1}^{N} [\text{log}(1+y_i)-\text{log}(1+\hat{y}_i)]^2,
\end{equation}
where $\hat{y}$ is the predicted value. Each model was trained for 100 epochs, and we choose a batch size of 100 as the most optimal choice. Tables III and IV in the supplementary material available at Ref.~\cite{ventagli_2024_11216586} summarize the most significant networks we tested. Each time we train a specific model, we do not fix the seed to which the network parameters have been randomly drawn. 
We can see that models perform better increasing the number of nodes, and, in general, considering dataset with larger values of noise injections $n$ and size of observation set $N$ also reduces further the loss. For the $M$-$R$ data, the best model we trained is composed by a hidden layers with 70 nodes for the dataset with $n=300$ and $N=30$ with a loss of $\sim 0.01799$; while for the $M$-$R$-$k_2$ the best loss we obtain is $\sim 0.01794$ for a single hidden layer network with 180 nodes, for the dataset corresponding to $n=300$ and $N=30$.

\begin{figure}[ht]
\centering
    \includegraphics[width=0.7\textwidth]{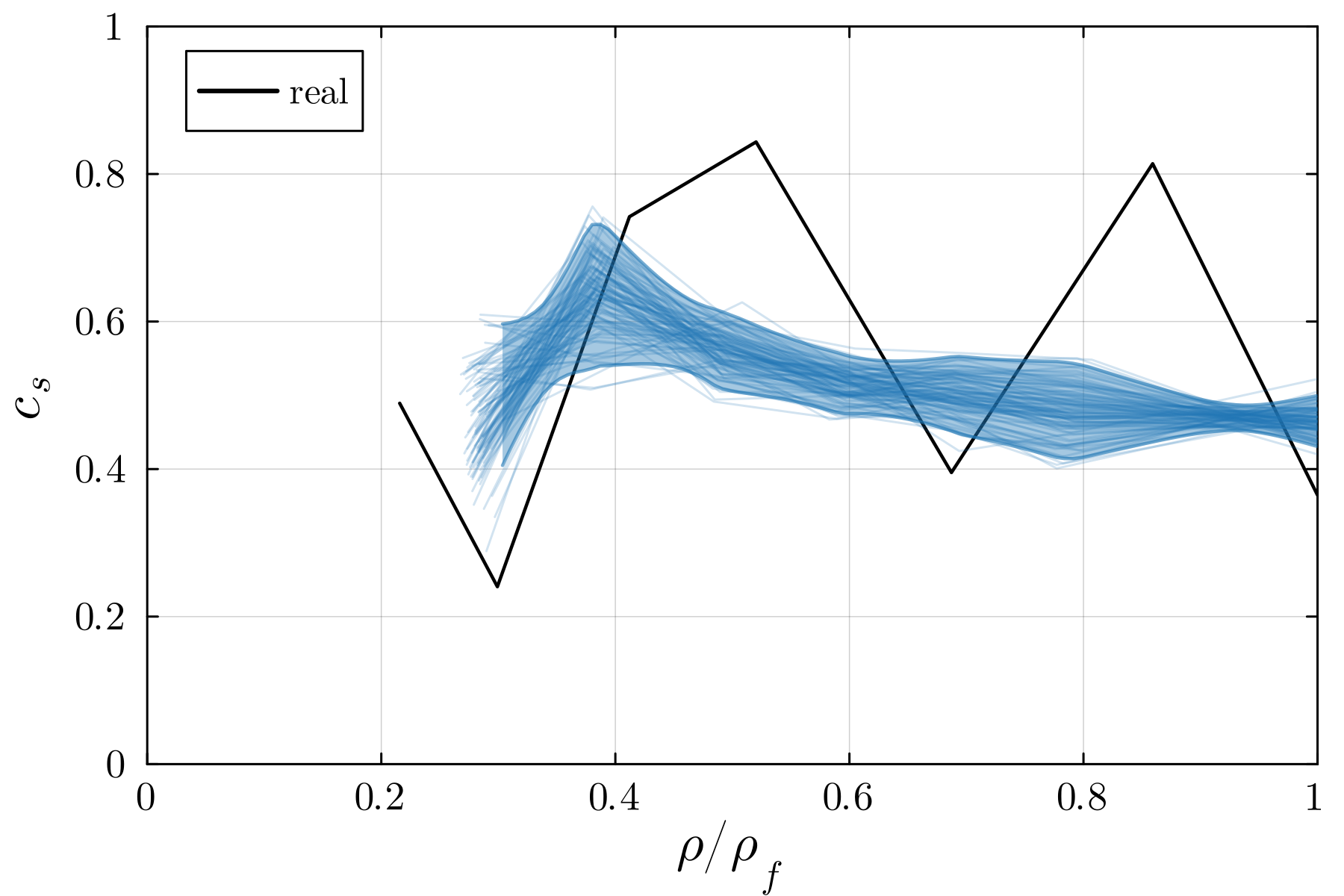}%
    \caption{A specific random sample for the speed-of-sound vs normalized mass-density parametrization, where $\rho_f=12\,\rho_0$ is the last fixed value for the mass density. The black line describes the real model, while blue curves represent all predicted profiles by the regression deep network applied on test dataset augmented by injecting Gaussian noise $n=100$ times, following the procedure described in Sec.~\ref{Sec:dataset}. We also show the $2\, \sigma$ confidence interval for the network predictions as a blue shaded region, obtained by computing the mean and the standard deviation of the predictions at each value of renormalized mass density. The network predictions prefers to `settle' around a mean value of $c_s$ such that $\langle c_s \rangle \sim 0.5$ for $\rho/\rho_f \to 1$. This behavior is retrieved generically, and is not related to the specific speed-of-sound model we show here.}
	\label{fig:csProfile}
\end{figure}

Let us now discuss the EOSs predicted from our deterministic network. We only present the results for the best performing model we trained for the $M$-$R$-$k_2$ dataset, i.e. we consider the network trained with $n=300$ and $N=30$, but the discussion holds more generically. To test our model we produce a 100 new random generated speed-of-sound models, thus obtaining $2 \times 10^3$ new EOSs, of which only $128$ survive the maximum mass test. For each EOS we generate $N=30$ triplets of $(M_i,R_i, k_{2,i})$ with either $n=10$ or $n=100$ noise injections each. Note that the accuracy for the classification model is $87.3\%$ and $87.0\%$ for the $n=10$ or $n=100$ test dataset respectively.

\begin{figure}[ht]
\centering
    \includegraphics[width=0.7\textwidth]{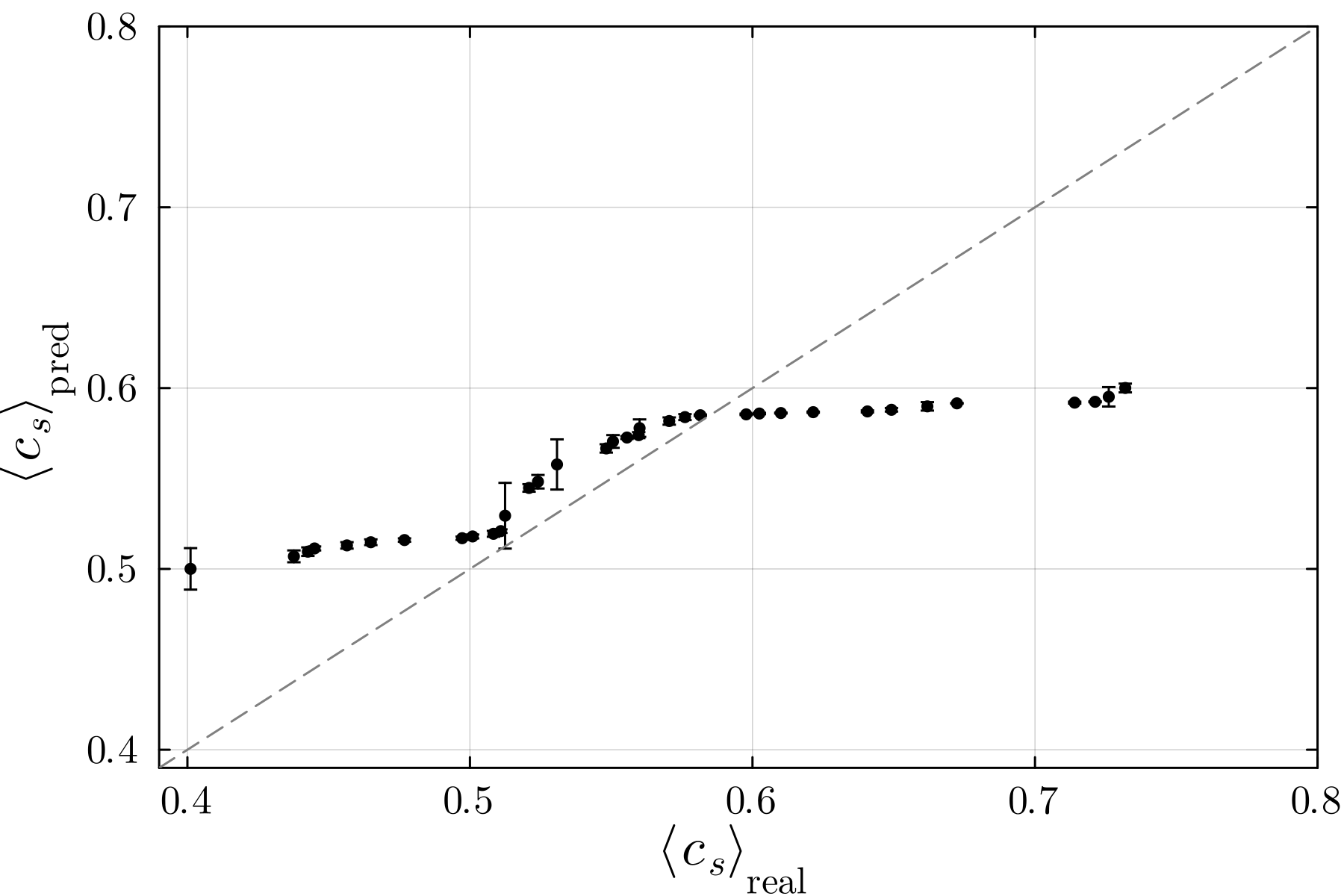}%
    \caption{Predicted mean speed-of-sound $\langle c_s \rangle$ with a $3\,\sigma$ confidence interval as a function of the ``real" (predicted from simulations) mean value for $c_s$ for all possible different speed-of-sound realizations in the test dataset. Almost all predicted values are not in agreement with the expected ones. The deep network tends to smooth out the sound speed profile, and misses large increases or decreases in the profile, which are responsible for oscillations on the mean value of $c_s$. Indeed, one can notice that the former are bound in the interval $(0.5,0.6)$, whereas for the latter one finds $\langle c_s \rangle_\text{real}\in (0.4,0.8)$. One can notice that deviations increase at the boundaries of such interval, where the mean value of the real speed-of-sound is either lower or higher to the predicted ones due to the presence of large oscillations in the real profile not reproduced by the network's realization.}
	\label{fig:csMean}
\end{figure}

In Fig.~\ref{fig:csProfile}, we show a specific random sample for the speed-of-sound parametrization (black solid line) and the predictions obtained from our network on the $n=100$ test dataset (blue solid lines). We also show the $2\,\sigma$ confidence interval for our predictions (blue shaded region). We can see that the network predictions are not able to fully cover the speed-of-sound profile, and tend to `settle' around a mean value of $c_s$ such that $\langle c_s \rangle \sim 0.5$ for $\rho/\rho_f \to 1$, where $\rho_f$ is the last value for the mass density, which we fix s.t. $\rho_f=12\,\rho_0$. We find this behavior to be generically true for all different speed-of-sound parametrizations. To better prove this point, in Fig.~\ref{fig:csMean} we show the predicted mean speed-of-sound $\langle c_s \rangle$ with a $3\,\sigma$ confidence interval as a function of the real one for all different speed-of-sound realizations. We notice that almost all predicted $\langle c_s \rangle$ are not in agreement with the expected values. In particular, all $\langle c_s \rangle_\text{pred}$ values are bound in the interval $(0.5,0.6)$, whereas the expected values cover a larger region, $\langle c_s \rangle_\text{real}\in (0.4,0.8)$. The deviations are more drastic at the boundaries of the latter interval. 

\begin{figure}[ht]
\centering
    \includegraphics[width=0.7\textwidth]{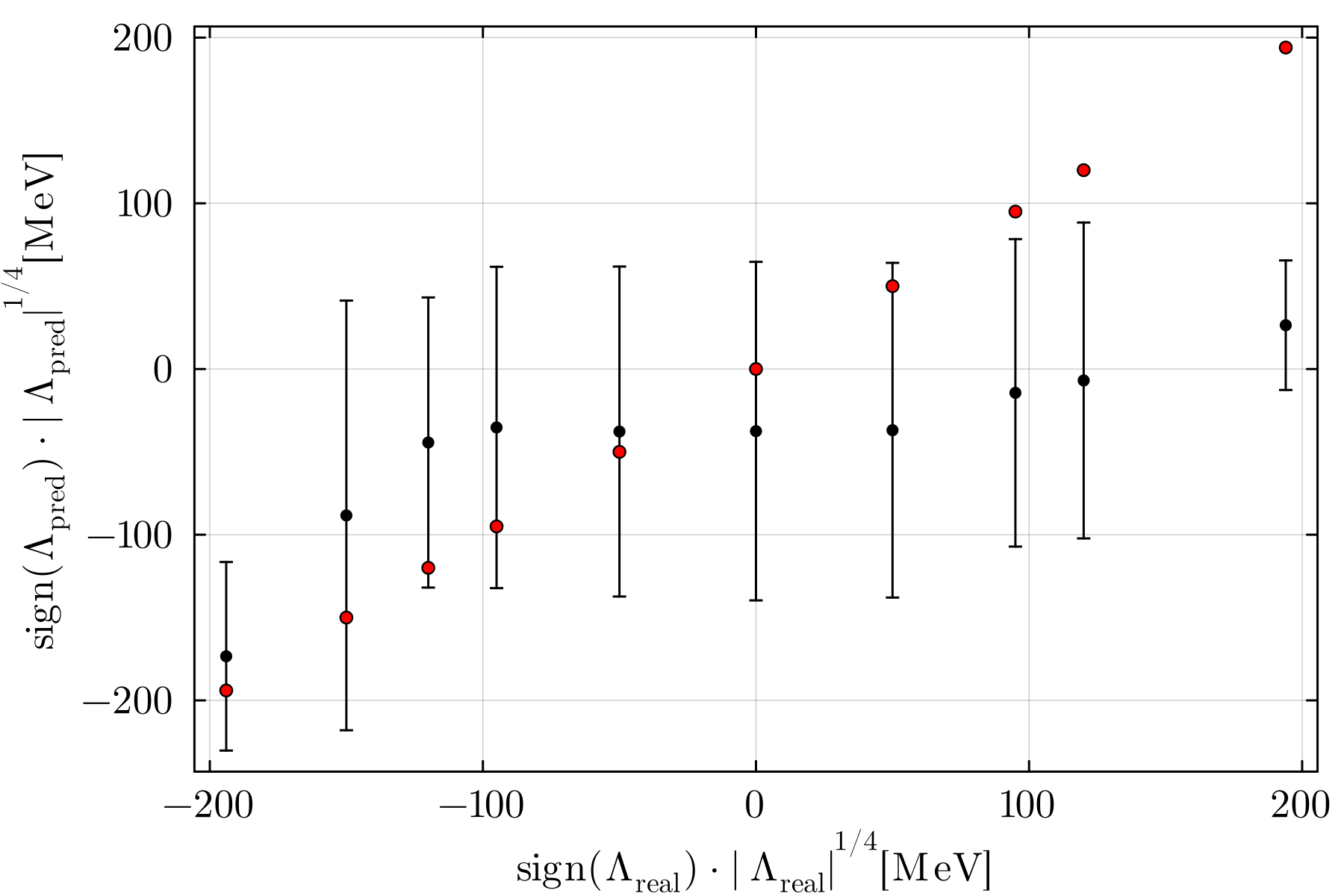}%
    \caption{Predicted mean vacuum energy shift with a $2\,\sigma$ confidence interval as a function of the real mean value for $\Lambda$ for all possible different values in the test dataset. Almost all predicted values lie within the $2\,\sigma$ intervals, with the exception of very large positive values of $\Lambda$. In particular, the highest possible value for the vacuum energy shift, i.e. $\Lambda=(194\,\text{MeV})^4$, is not correctly predicted, even when considering a $3\,\sigma$ interval. The reason why our network can better identify a large negative vacuum energy, rather than a (large) positive one, lies on the different properties of these two cases, which can be considered as two distinct family of EOSs: configurations that admits a wide range of values for $\Lambda$ and those that are allowed only if a large negative jump in vacuum energy is introduced. This leads to different stellar properties. In particular, the second family displays peculiar $M$-$R$ relations, as shown in Fig.~\ref{fig:MR}, and the network can better distinguish them between all possible configurations. Moreover, our training dataset is biased towards lower values of the vacuum energy shift, since these values yield a larger set of physically acceptable solutions of the gravity equations.}\label{fig:LambdaMean}
\end{figure}

Let us now focus on the results for the predicted vacuum energy shift. In Fig.~\ref{fig:LambdaMean}, we show the predicted mean $\Lambda$ with a $2\,\sigma$ confidence interval as a function of the real one for all $10$ possible values considered. We considered here the $n=100$ noise injections dataset. While the network can successfully predict a very large negative value of $\Lambda$, it tends to perform poorly as $\Lambda$ increases. While most of predicted values lie within the $2\,\sigma$ intervals of the ``real'' (predicted by simulations) ones, the measures are also consistent with $\Lambda=0$. In particular, we notice that the highest possible value for the vacuum energy shift, i.e. $\Lambda=(194\,\text{MeV})^4$, is not correctly predicted, even when considering a $3\,\sigma$ interval. We expect that the reason why our networks can better identify a large negative phase transition, rather than a (large) positive one, lies on the different properties of these two cases. EOSs with a vacuum energy phase transition can be divided into two classes~\cite{Ventagli:2024cho}: EOSs that admits a wide range of values for $\Lambda$ and EOSs that are allowed only if a large negative jump in vacuum energy is introduced. These lead to different $M$-$R$ and $k_2$-$R$ relation properties. In particular, EOSs that admit a large negative $\Lambda$ display peculiar $M$-$R$ relations and the network can better distinguish them between all possible configurations. Moreover, our training dataset is biased towards lower values of the vacuum energy shift, since these values yield a larger set of physically acceptable solutions of the gravity equations. As a result, our network cannot confidently constrain the value of the jump in vacuum energy for the first family of EOSs.

\begin{figure*}[hbt!]
\begin{center}
\includegraphics[width=0.5\linewidth]{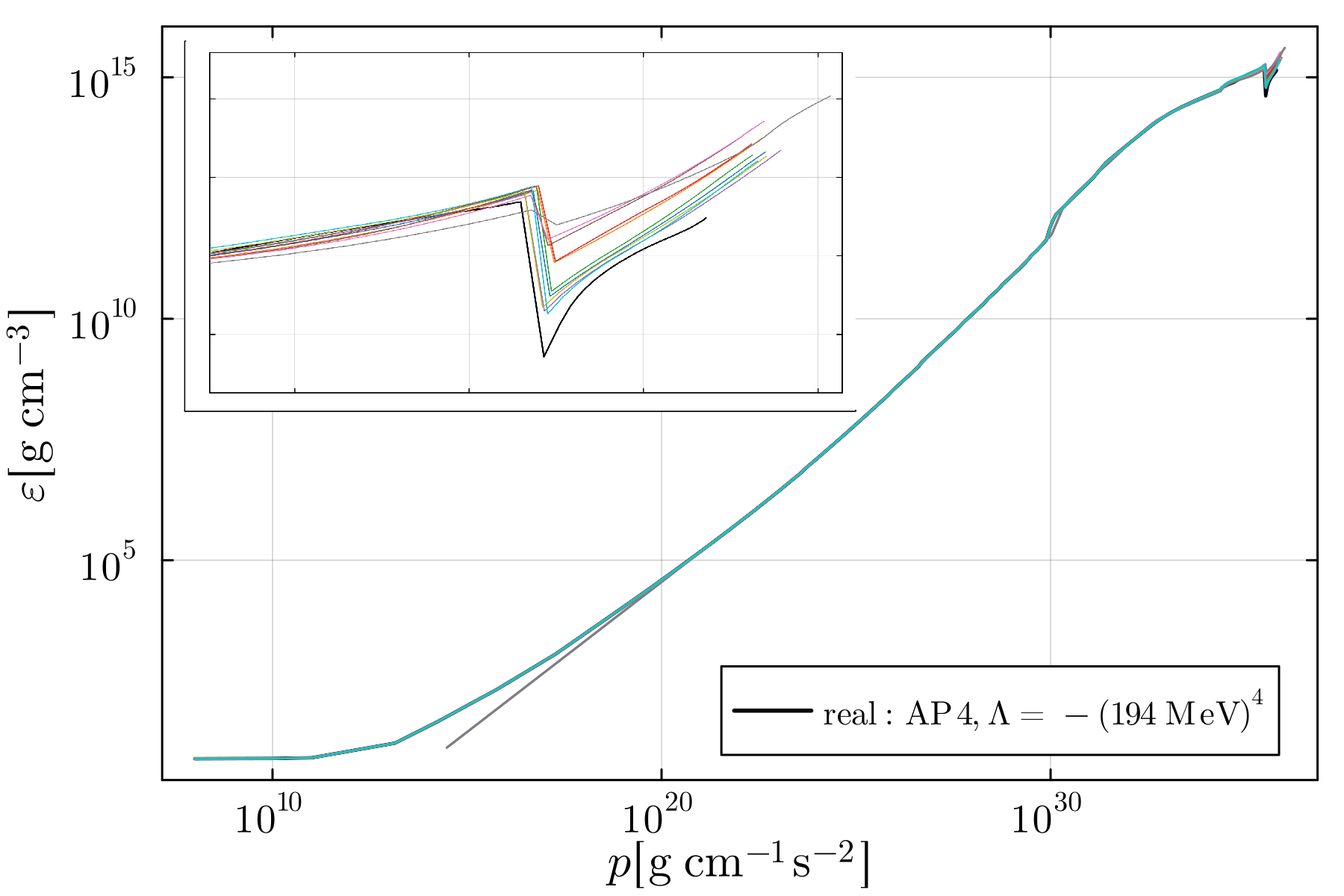}\hfill
\includegraphics[width=0.5\linewidth]{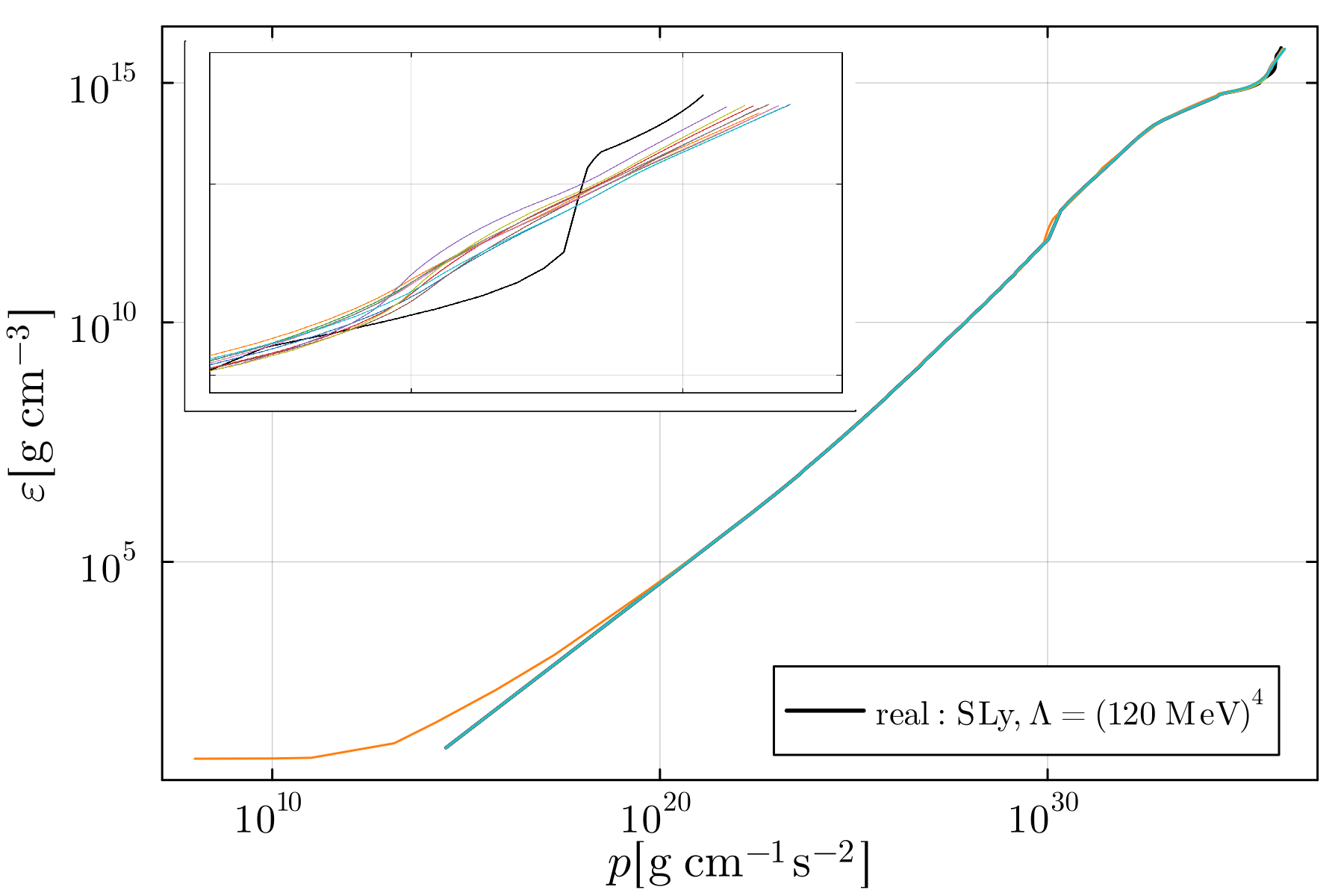}
\caption{Two specific EOS models (black lines) and their predictions for the test dataset augmented by injecting Gaussian noise $n=10$ times, following the procedure described in Sec.~\ref{Sec:dataset}. On the left (right) panel we show an EOS built from the AP4 (SLy) and with a negative (positive) jump in vacuum energy, with $\Lambda=-(194\, \text{MeV})^4$ ($\Lambda=(120\, \text{MeV})^4$). The two cases also use different speed-of-sound parametrizations. In almost all cases, the realizations have been predicted with the correct low-density model. At higher density scales, for the case of a negative vacuum energy transition, all predictions correctly display a sudden decrease in energy density. For a positive jump in $\Lambda$, however, the typical phase transition behavior is lost, and the energy density only slowly increases. The reason why our network can better identify a large negative vacuum energy, rather than a (large) positive one, lies on the different properties of these two cases. As it was pointed out in Ref.~\cite{Ventagli:2024cho}, there are two families of EOSs: those that admits a wide range of values for $\Lambda$ and those that are allowed only if a large negative jump in vacuum energy is introduced. This leads to different stellar properties. In particular, the second family displays peculiar $M$-$R$ relations and the network can better distinguish them between all possible configurations. Note also that our training dataset is biased towards lower values of the vacuum energy jump, since these choices lead to a larger set of physically acceptable solutions of the gravity equations.}
	\label{fig:EOS}
\end{center}
\end{figure*}

We now focus on the results coming from both the classification and the regression networks. This allows us to construct the full EOS, composed by the three regions described in section~\ref{Sec:NS}. In Fig.~\ref{fig:EOS}, we show two specific EOS models (black line) and their predictions for the $n=10$ test dataset. Note that the two expected EOSs correspond to different low density descriptions, different speed-of-sound parametrization and different vacuum energy shifts. On the left panel we present an EOS built from the AP4 and with a negative jump in vacuum energy, with $\Lambda=-(194\, \text{MeV})^4$. For the low density region, $8$ out of $10$ realizations have been predicted with the correct model (AP4). Moreover, at higher density scales all predictions correctly follow the decrease in energy density due to the vacuum energy shifts. On the right panel we show the case of an EOS built from the SLy model with a positive vacuum energy jump, with $\Lambda=(120\, \text{MeV})^4$. In this instance, only a single realization is predicted with the wrong low-density description. However, at high densities, the predictions follow the expected behavior less rigorously, in particular the increase in energy density caused by the positive jump in vacuum energy is less abrupt, and one loses the typical phase transition behavior. Once again, the reason why our networks can better identify a large negative phase transition, rather than a (large) positive one, lies on our training dataset being biased towards lower values of the vacuum energy shift and on the fact that these two cases belong to different EOS' families, which leads to different stellar properties~\cite{Ventagli:2024cho}.

\begin{figure*}[hbt!]
\begin{center}
\includegraphics[width=0.5\linewidth]{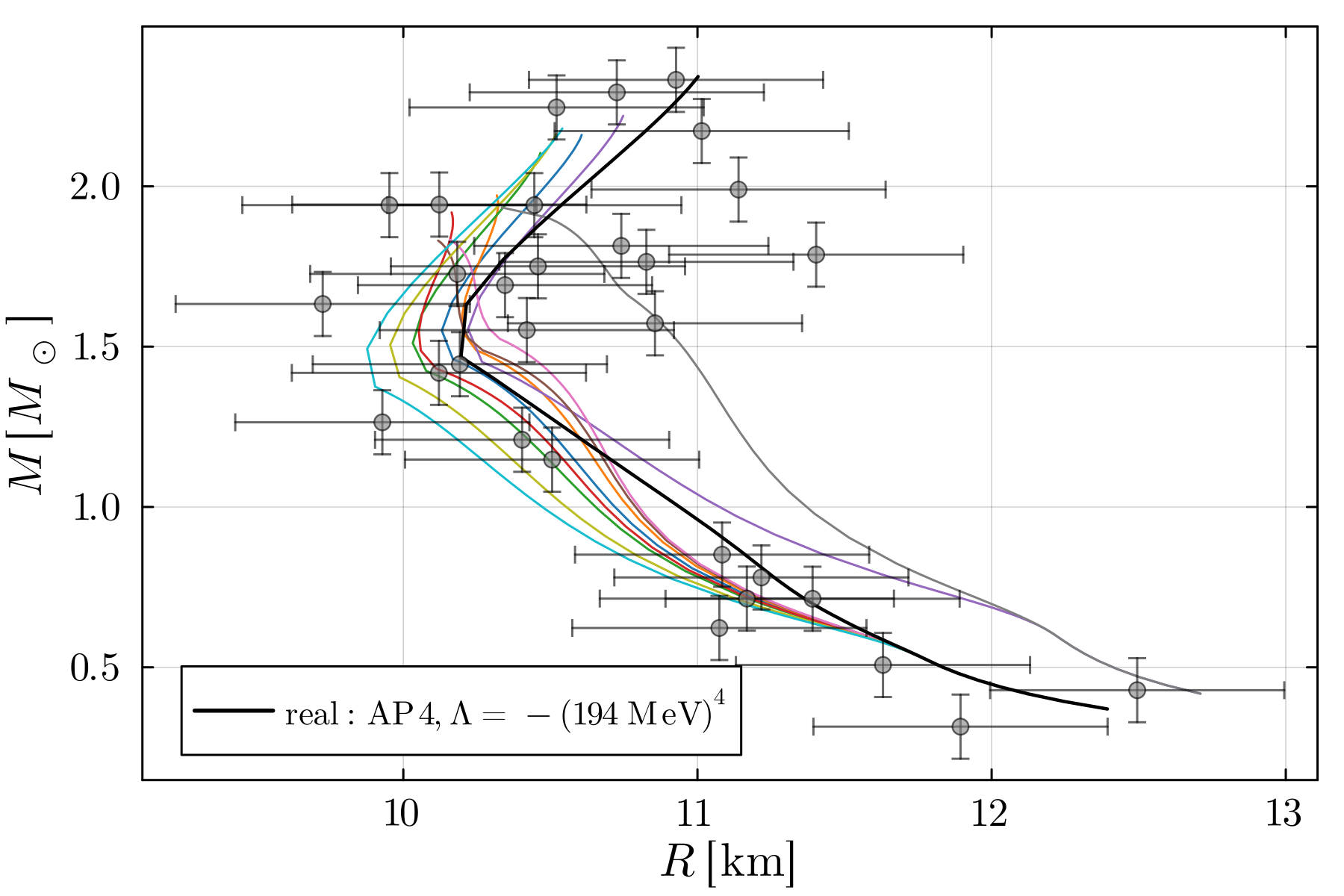}\hfill
\includegraphics[width=0.5\linewidth]{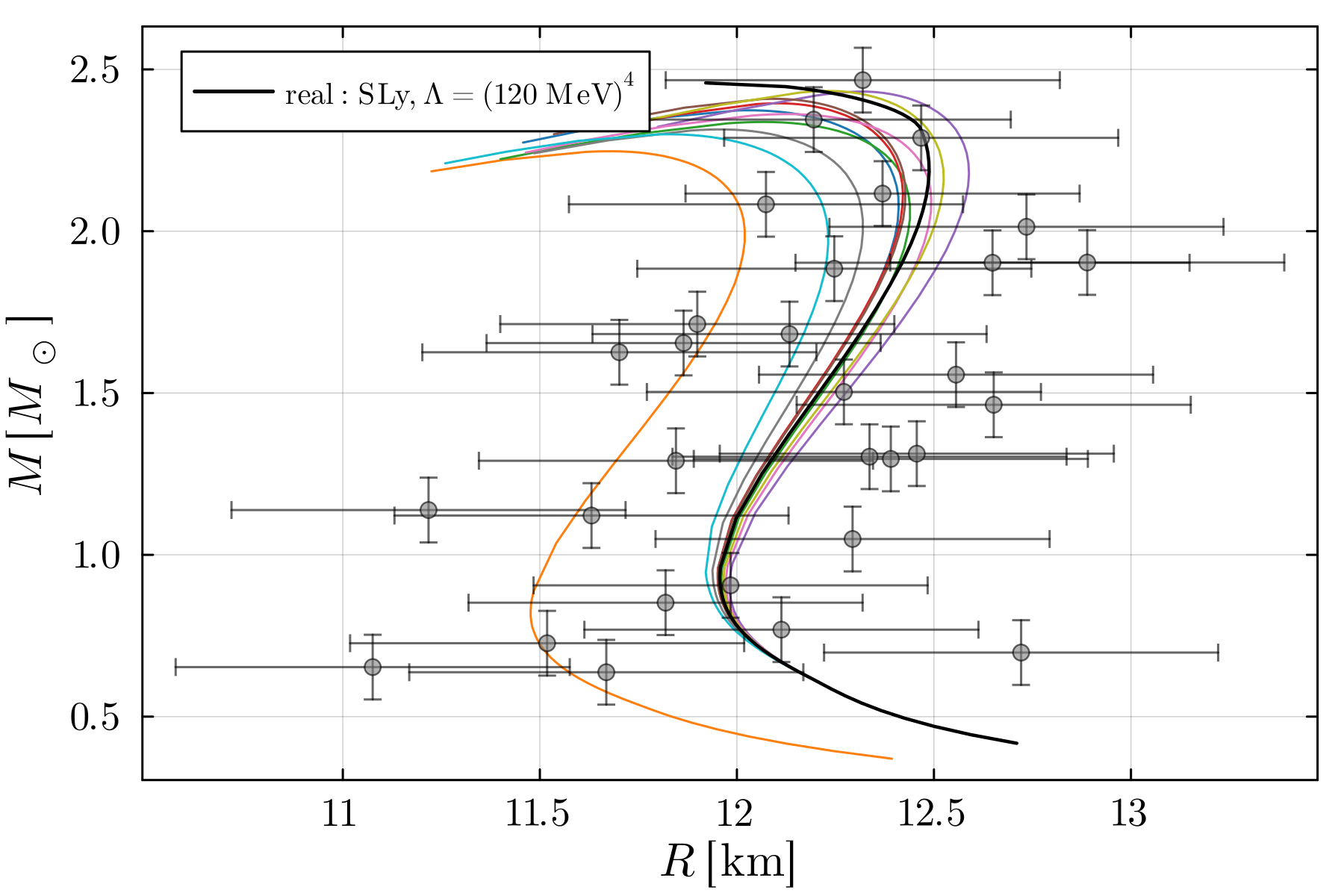}
\caption{Mass-radius relations for two specific EOS models, corresponding to those presented in Fig.~\ref{fig:EOS}. On the left (right) panel we show an EOS built from the AP4 (SLy) and with a negative (positive) jump in vacuum energy, with $\Lambda=-(194\, \text{MeV})^4$ ($\Lambda=(120\, \text{MeV})^4$). Curves obtained from solving the TOV equations for the real EOS are shown in black, whereas the network predictions are shown with different colors, maintaining the same mapping as in Fig.~\ref{fig:EOS}. We also include, as an example, a single set of observation points with injected Gaussian noise as described with a $1\, \sigma$ confidence interval, as described in Sec.~\ref{Sec:dataset}, used as an input for the network. All of the observation points are within $3\, \sigma$ from the $M$-$R$ curves obtained from the predicted realizations.}
	\label{fig:MR}
\end{center}
\end{figure*}

This can be seen in Fig.~\ref{fig:MR}, where we show the $M$-$R$ relations obtained by solving the TOV equations for both the expected EOS (black line) and the predicted ones for the two EOS models considered in Fig.~\ref{fig:EOS}. Note that the color mapping is maintained, i.e. an EOS described by a specific color line in Fig.~\ref{fig:EOS} yields a $M$-$R$ curve of the same color in Fig.~\ref{fig:MR}. We also include, as an example, a single set of observation points $O=\{M_i,R_i\}$ with injected Gaussian noise as described in Eq.~\eqref{eq:injnoise} with a $1\, \sigma$ confidence interval. For both cases presented, all of the observation points, which we remind are used as input for our networks, are within $3\, \sigma$ from the $M$-$R$ curves obtained from the predicted realizations. This is true even when the model predicts a wrong low-density model. 

\begin{figure}[hbt!]
\begin{center}
 \includegraphics[width=0.5\textwidth]{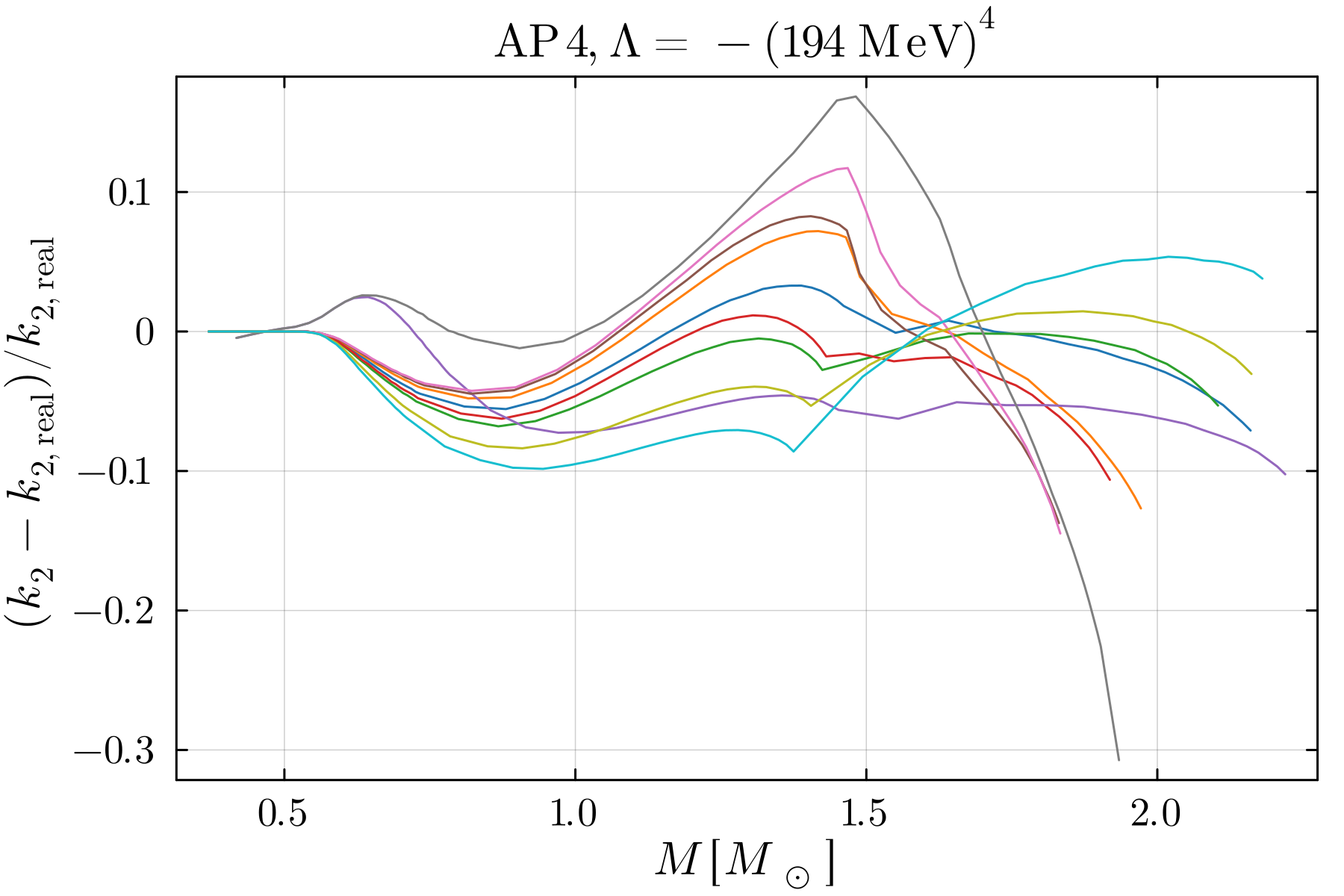}%
\includegraphics[width=0.5\textwidth]{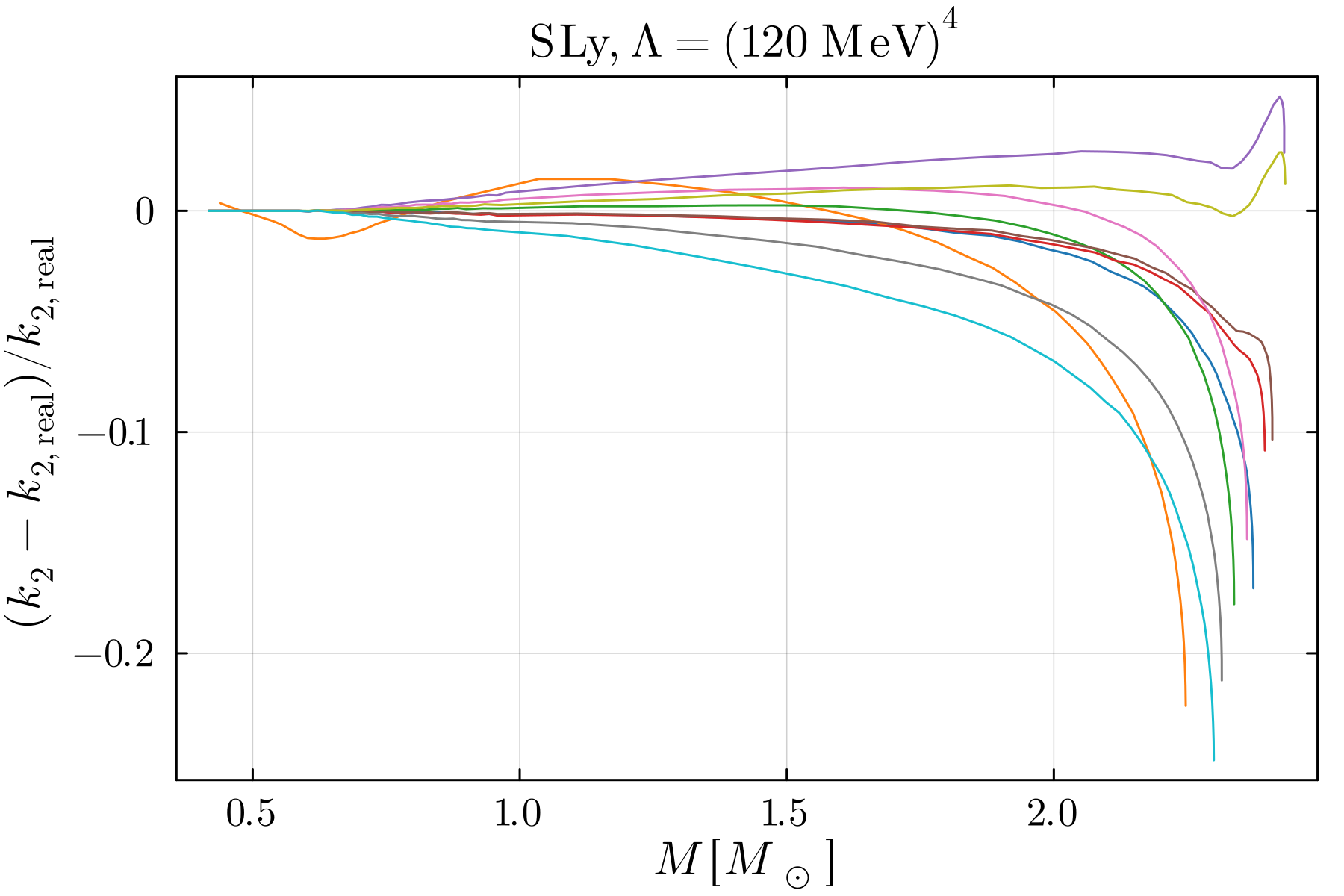}%
\caption{Fractional error for the tidal Love number as a function of the total stellar mass for two specific EOS models, corresponding to those presented in Fig.~\ref{fig:EOS}. On the right (left) panel we show an EOS built from the AP4 (SLy) and with a negative (positive) jump in vacuum energy, with $\Lambda=-(194\, \text{MeV})^4$ ($\Lambda=(120\, \text{MeV})^4$). We maintain the color mapping as in Fig.~\ref{fig:EOS}. The fractional error is typically around $10\%$ at intermediate masses, while it can increase up to $30\%$ in the case of more massive stars. This is to be expected since more massive configurations have higher central pressure, where the EOSs predictions deviates more significantly from the expected one, as can be seen from Fig.~\ref{fig:EOS}.}
	\label{fig:k2Mfrac}
\end{center}
\end{figure}

Finally, we discuss the $k_2$-$M$ relations obtained by solving Eq.~\ref{eq:diffH} for the expected EOS (black line) and those obtained from our network. Once again, we maintain the color mapping as in Fig.~\ref{fig:EOS}. We do not include a set of observation points, to not complicate further the figure, however all mock observations are in agreement with the results produced from the predicted realizations. In Fig.~\ref{fig:k2Mfrac}, we show the fractional error for the tidal Love number $(k_2 - k_{2,\,\text{real}})/k_{2,\,\text{real}}$. We can see that, typically, this is around $10\%$ at intermediate masses, while it can increase up to $30\%$ in the case of more massive stars.

\begin{figure*}[hbt!]
\begin{center}
\includegraphics[width=0.5\linewidth]{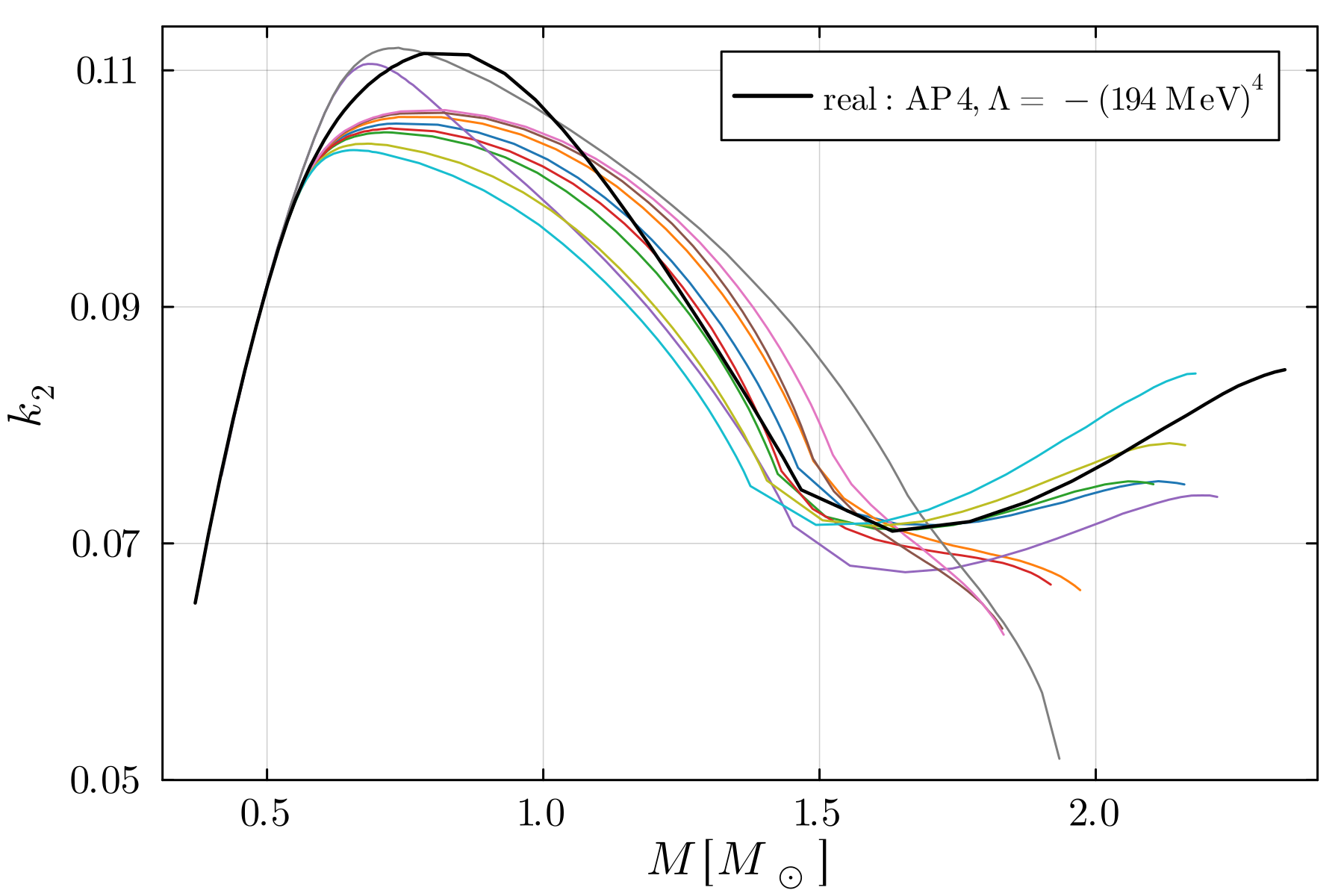}\hfill
\includegraphics[width=0.5\linewidth]{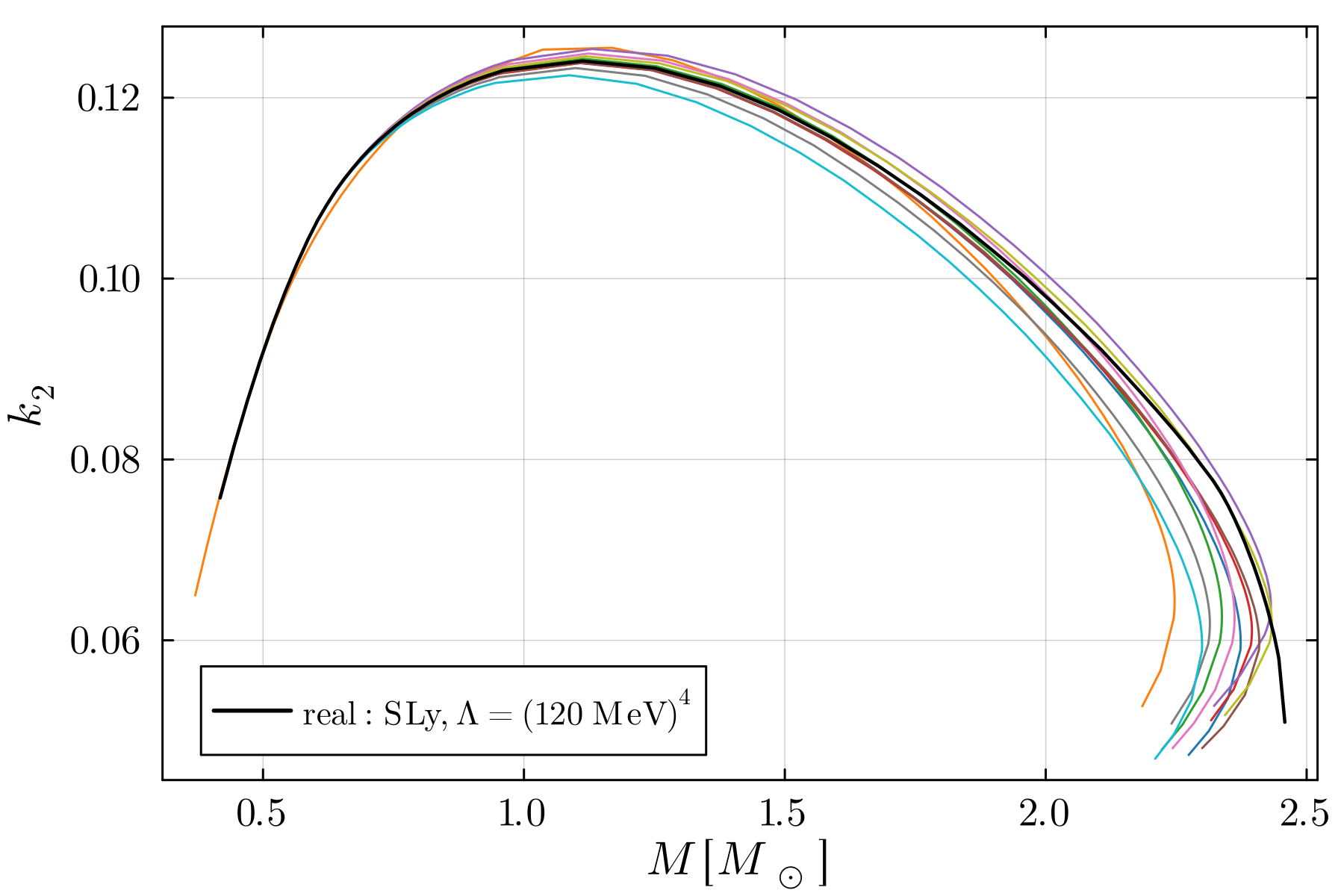}
\caption{Tidal Love number-mass relations for two specific EOS models, corresponding to those presented in Fig.~\ref{fig:EOS}. On the left (right) panel we show an EOS built from the AP4 (SLy) and with a negative (positive) jump in vacuum energy, with $\Lambda=-(194\, \text{MeV})^4$ ($\Lambda=(120\, \text{MeV})^4$). Curves obtained from solving the TOV equations for the real EOS are shown in black, whereas the network predictions are shown with different colors, maintaining the same mapping as in Fig.~\ref{fig:EOS}. We do not include a set of observation points, to not complicate further the figure, however all input observations are in agreement with the results produced from the predicted realizations.}
	\label{fig:k2M}
\end{center}
\end{figure*}

\subsection{Bayesian deep network}\label{Sec:Bayesian}


\begin{figure}[hbt!] 
\begin{center}
  \includegraphics[width=0.49\textwidth]{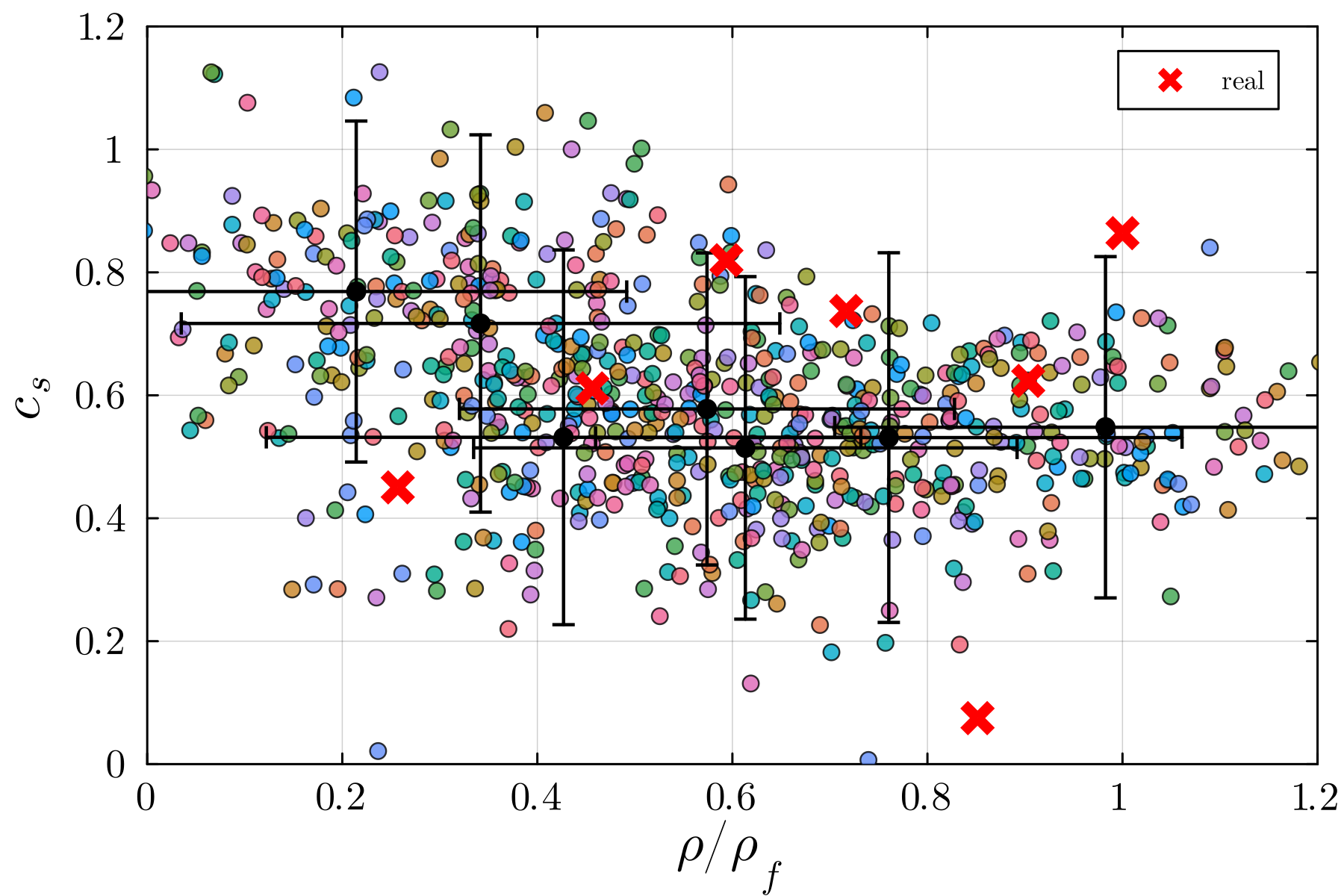}
  \includegraphics[width=0.49\textwidth]{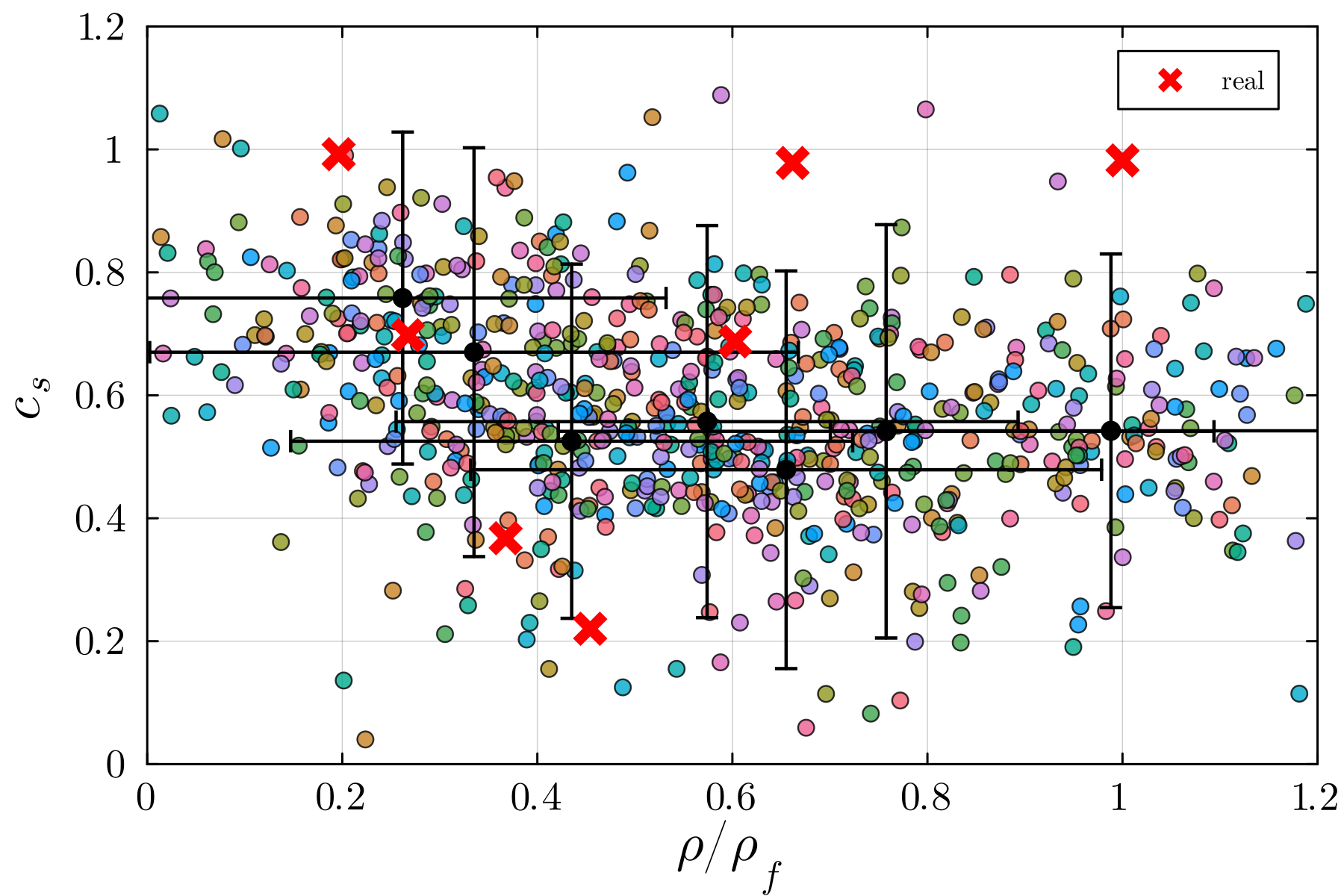}
  \caption{A statistical sample of the predictions (features) as outputted from our Bayesian (probabilistic) deep network, and for two input choices according to our dataset's numbering, corresponding to the $100$-th (right) and $1000$-th (left) input data point respectively. For a given input of masses, radii and tidal Love numbers, or equivalently a given input of the EOS, the output of the deep network corresponds to a particular realisation of the $7$ pairs of mass density-sound speed. We use the dataset where each input data point comprises of $30$ triplets of total mass, radius, tidal number, and we consider an $n_{\text{samples}} = 100$ realisations of the output (coloured points). The red crosses denote the expected (``real") values on the mass density-sound speed plane for the given EOS. The black dots and the associated error-bars are computed as the average and $2\sigma$ errors respectively, given the statistical sample for each column of the output's array. As long as the predictions (red crosses) lie within the respective $2\sigma$ error bar, the prediction is considered successful, according to our evaluation pipeline, explained in Sec. \ref{Sec:Bayesian}. Notice that in the plots we chose not to discard predictions which are outside the physical range for the mass and sound speed we consider, i.e points lying outside the range $[0,1]$. Though individual predictions may lie outside this range, central values and respective error bars are consistent with our assumed physical range of values. It is also to be noticed how the size of error bars change with the assumed EOS provided as input to the deep network.} \label{fig:cs-prediction}
  \end{center}
\end{figure}

\begin{figure}[!tbp]
\begin{center}
  \subfloat[Histogram for mass density values.]{\includegraphics[width=0.5\linewidth]{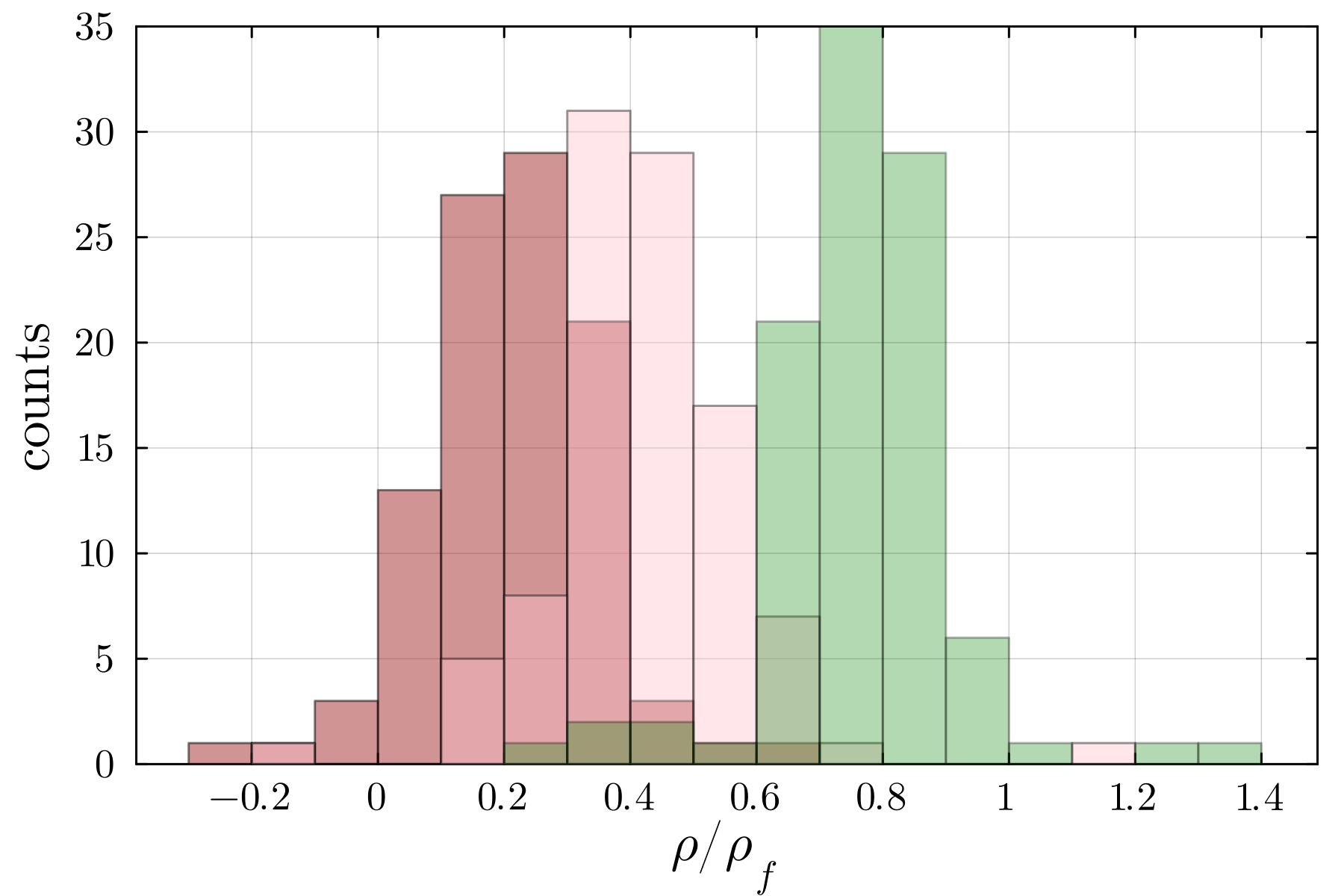}}\hfill
  \subfloat[Histogram for sound speed values.] {\includegraphics[width=0.5\linewidth]{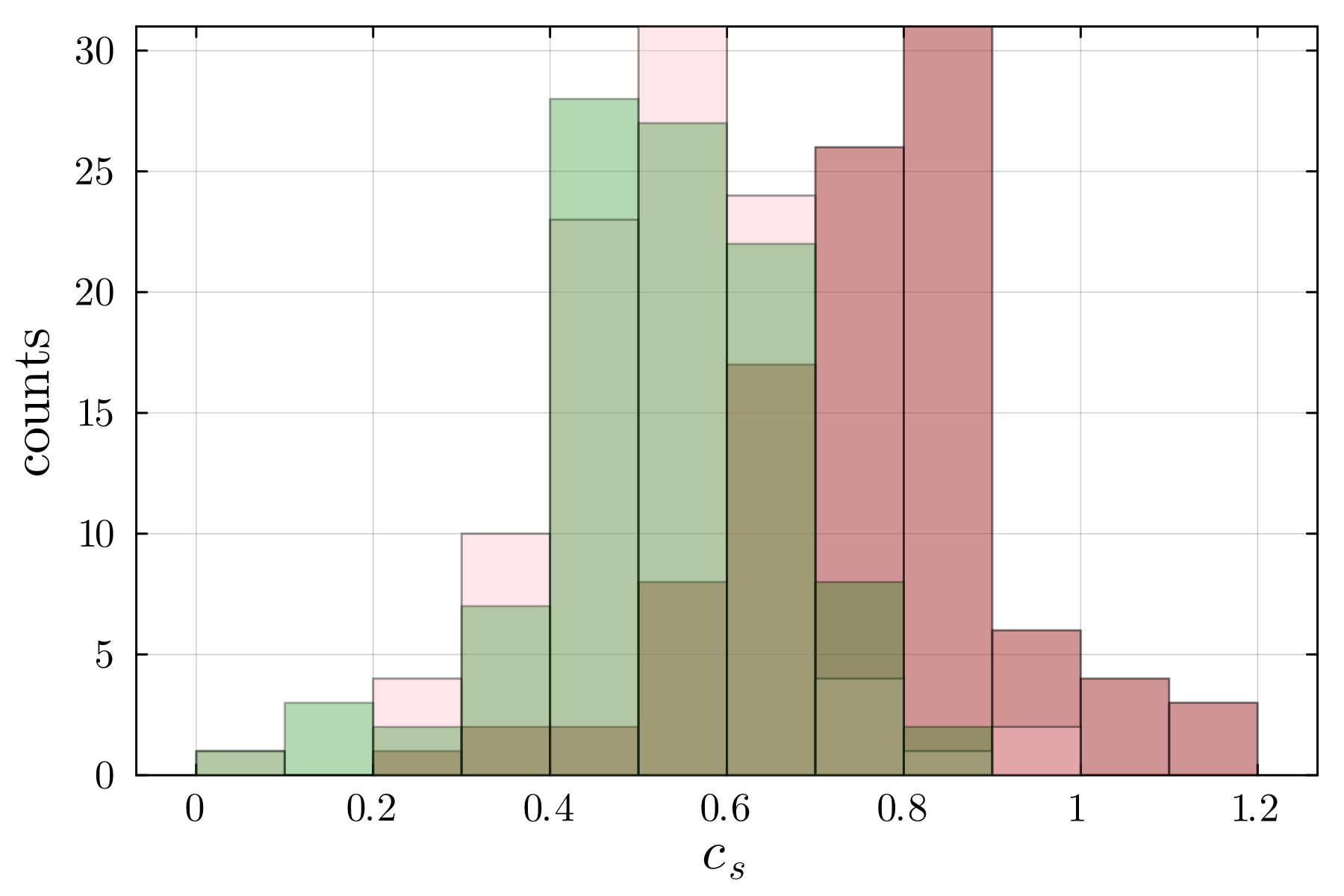}}
  \caption{Histograms computed from the output of our Bayesian (probabilistic) deep network, for the EOS choice $24601$, which corresponds to half of the total size of the testing set. They describe the counts of the distribution for the mass density and sound speed for three different columns of the network's output, namely columns $1,3,6$ from left to right in the first plot. We remind that given an input of masses, radii and tidal Love numbers, or equivalently a given input of the EOS, the output of the deep network corresponds to $7$ values of mass density and $7$ values for the sound speed respectively. The sample in above histograms was computed considering $100$ realisations of the network's output. One notices that the histograms for the sound speed are more concentrated around a particular value close to $0.5$, whereas the histograms for the mass density cover a broad range of masses. From such histograms we compute the mean values and the corresponding standard deviation as shown in Fig. \ref{fig:cs-prediction}.} \label{fig:mass-cs-histograms}
  \end{center}
\end{figure}

\begin{figure}[!tbp]
\begin{center}
  \subfloat['True' vacuum energy value $\simeq -(150\, \text{MeV})^4$.]{\includegraphics[width=0.5\linewidth]{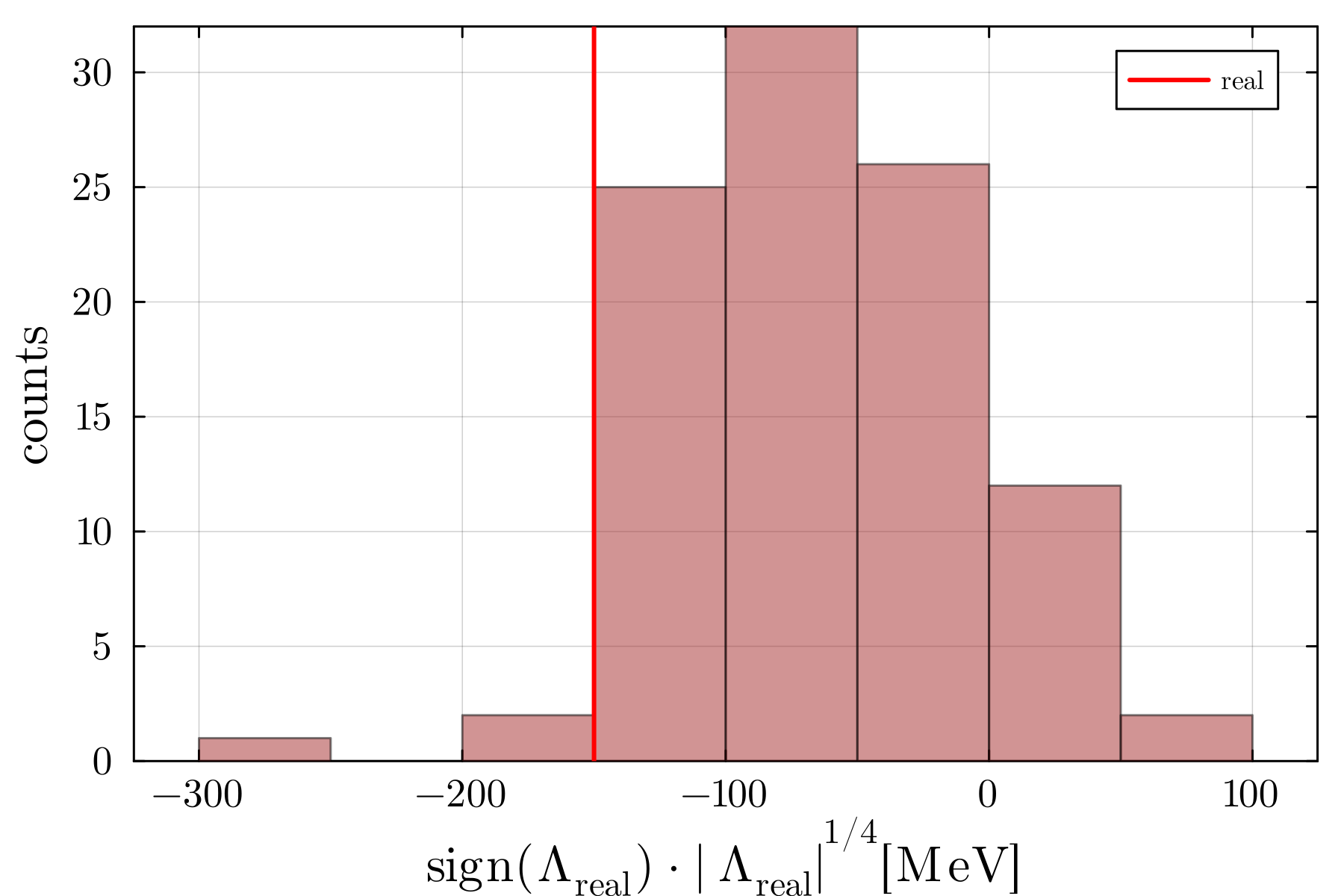}}\hfill
  \subfloat['True' vacuum energy value $\simeq (194\,\text{MeV})^4$.] {\includegraphics[width=0.5\linewidth]{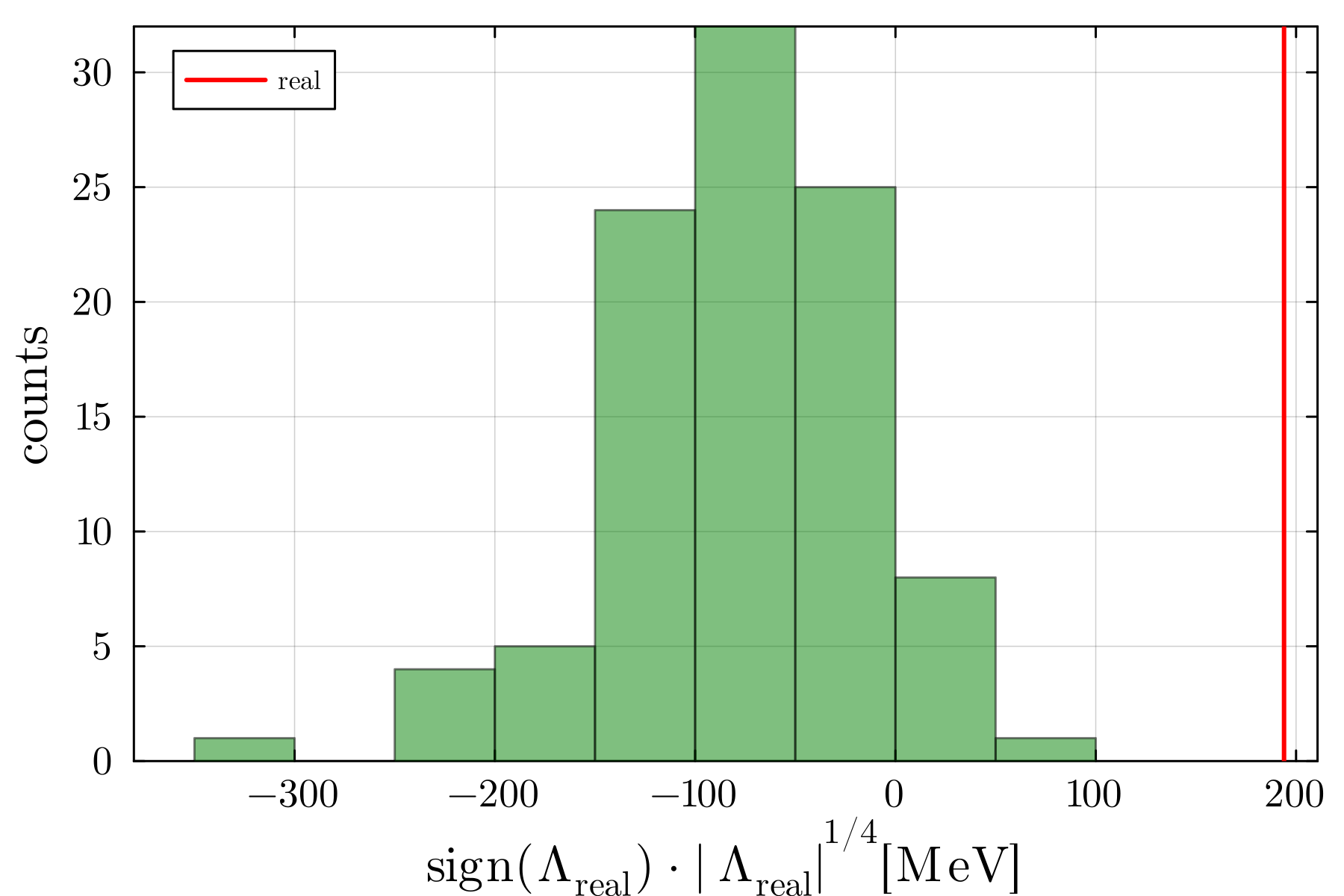}}
  \caption{Histograms based on the prediction of our Bayesian (probabilistic) deep network for the vacuum energy associated with the phase transition in the star, based on a statistical realisation for each input. The two inputs correspond to two different ``true" values of the vacuum energy. The performance of the network is not as satisfactory as with the similar predictions of the sound speed value, evident in the wider spread of the histograms. We notice that the network appears to prefer a central value of about $\simeq -(70\, \text{MeV})^4$  for both inputs, while the standard deviation is approximately $\simeq (130\,\text{MeV})^4$ for both cases. The network performs better with lower vacuum values - in the upper plot, the network predicts the ``true" value within $2\sigma$, whereas in the second case in more than $4\sigma$. }\label{fig:Lambda-histograms}
  \end{center}
\end{figure}

In the deep network discussed earlier, the weights and biases were point estimates computed during the training process. However, in order to capture the statistical nature of the data and the epistemic errors in them, one needs to resort to a statistical approach. In the probabilistic, or Bayesian view of deep networks, the weights and biases at each node are not point estimates any more, but they are drawn from certain probability density functions (p.d.f's), which aim to describe the statistical uncertainties in the input data. At the heart of this approach lies {\it Bayes' theorem},
 \begin{equation} 
 p({\boldsymbol{\theta}}|D) = \frac{p(\boldsymbol{\theta})\cdot p(D| \boldsymbol{\theta})}{\sum_{\boldsymbol{\theta}} p(D|\boldsymbol{\theta}) \cdot p(\boldsymbol{\theta})} \equiv  \frac{p(\boldsymbol{\theta}, D)}{p(D)} ,
 \end{equation}
 where $D$ is the set of input data (observations), and $\boldsymbol{\theta}$ is a vector of random variables (parameters) we are trying to fit, and here we will assume it is discrete. A direct evaluation of the posterior $p({\boldsymbol{\theta}}|D)$ is a tedious task, and one typically needs to resort to approximations to compute it. Here, we will use the approach of {\it variational Bayes} which allows for an approximate computation of the posterior. In this regard, we start with an ansatz, or ``surrogate" posterior which approximates the true one as
 \begin{equation}
 q({\boldsymbol{\theta}}) \approx p({\boldsymbol{\theta}}|D).
 \end{equation}
The goal of variational Bayesian inference is to re-construct the posterior $q$ so that the difference with the true posterior, $p({\boldsymbol{\theta}}|D)$, is minimal. The difference between the p.d.f's is measured through the so--called Kullback-Leibler (KL) divergence. Given two p.d.f's of a random variable $\boldsymbol{\theta}$, $p(\boldsymbol{\theta})$ and $q(\boldsymbol{\theta})$ respectively, the KL divergence is defined as,
\begin{equation} \label{eq:KL}
\text{KL}\left( q || p \right) \equiv \sum_{\boldsymbol{\theta}} q(\boldsymbol{\theta}) \log \frac{q(\boldsymbol{\theta})}{p(\boldsymbol{\theta} | D)}.
\end{equation}
In our approach, the approximate posterior, $q(\boldsymbol{\theta})$, will be based on some analytical, a priori ansatz, which we will compute after minimisation of the KL divergence \eqref{eq:KL}. The resulting posterior function will then be a proxy to the true one in Bayes' rule \footnote{We notice here that the KL divergence extends to the case of a continuous random variable through the replacement of the sum with an integral.}. It can be shown that minimising the KL divergence amounts to maximising the quantity
\begin{equation}
 \mathds{E}_{q} \left[  \log p(\boldsymbol{\theta}, D) - \log q(\boldsymbol{\theta}) \right] = \log p(\boldsymbol{\theta}) -\text{KL}\left( q || p \right),
\end{equation}
with $ \mathds{E}_{q}$ the expectation value with respect to the approximate posterior $q(\boldsymbol{\theta})$, i.e $ \mathds{E}_{q} [\ldots] \equiv \sum_{\boldsymbol{\theta}} q(\boldsymbol{\theta}) \left[ \ldots \right]$. This process results to the ``closest" posterior to the true one, under the given assumptions.

In the context of our deep network, our goal is to use the above procedure in order to model the posterior function $p(\boldsymbol{\theta} | D)$ for the weights and biases in each node of our deep network, so that the epistemic uncertainty in our data is accounted for. Therefore, the vector of parameters $\boldsymbol{\theta}$ corresponds to the weights and biases at each node. For the prior function $p(\boldsymbol{\theta})$ at each node we choose a diagonal, multi-dimensional Gaussian, with $\sigma = 1$ and zero covariance. For the approximate posterior $q(\boldsymbol{\theta})$ we choose a multi-dimensional, normal Gaussian with non-zero covariance. Therefore, the weights and biases at each node are initially uncorrelated, however, as the deep network gets trained, the dependence of each weight/bias to the other is modelled by the non-zero covariance of the posterior. The assumption of a functional form for the posterior can be certainly challenged, however, this choice makes our numerical construction more tractable, compared to a non-parametric reconstruction. We notice that we allow only for {\it epistemic uncertainty}, that is, the statistical uncertainty related to our data set. The conceptually different type of uncertainty, namely the aleatoric uncertainty, relates to inherently random effects. This uncertainty could be in principle implemented in a straight-forward manner, but here we are interested only in the systematic effects in the analysis of our data.
 
We implement our Bayesian deep network on {\it TensorFlow} making use of the class of {\tt DenseVariational} layers which is part of the {\tt tensorflow\_probability} library. These are the probabilistic extension of the more standard {\tt Dense} layers. We choose a {\tt LeakyReLU} for our activation function and the mean-square error for our loss function respectively. The activation function is applied to all hidden layers, but not in the output layer - we comment further below about this choice. The minimisation of the loss function is performed through the {\tt RMSprop} algorithm, which shows good-enough convergence at training. We found that a network with $4$ hidden layers, each with $16$ units, was a sufficiently minimal choice yielding satisfactory results for our purposes. Notice that the input and output of the network remained the same as with the standard (non-probabilistic) version of the deep network. 

In order to evaluate the network we follow the following procedure. For a given input (e.g. $7$ pairs of total mass-radius values and the value of the vacuum energy shift), we execute the network iteratively in order to produce different realisations of the output data, that is the value of the vacuum energy shift and $7$ pairs of mass density and sound speed. Focusing on the predictions of the latter, for $N$ different realisations we will have $7 \times N$ pairs of mass density and sound speed values, i.e $\{ ( \rho^{(1)}_1, c_{s\, 1}^{(1)}), \ldots, ( \rho^{(7)}_1, c_{s\, 1}^{(7)}) \}$. Given the sample, we compute the mean quantities for each column 
\begin{equation}
\bar{\rho}^{(J)} = \frac{1}{N} \sum_{i=1}^{i = N} \rho_{i}^{(J)}, \; \; \; \; \bar{c}_{s}^{(J)} = \frac{1}{N}\sum_{i=1}^{i = N} c_{s\, i}^{(J)}, \; \; \; J = 1, \ldots, 7,
\end{equation}
where the index $i$ clearly counts the realisation. What is more, for each sub-sample labelled by the index $J$ we compute the corresponding standard deviation, $\sigma_{\rho}^{(J)}$, $\sigma_{c_s}^{(J)}$. We end up with a set of mean pairs of values  $\{ ( \bar{\rho}^{(1)}, \bar{c}_{s}^{(1)}), \ldots, ( \bar{\rho}^{(7)}, \bar{c}_{s}^{(7)})\}$ and their respective $\sigma$-errors $\{ ( \sigma_{\rho}^{(1)}, \sigma_{c_{s}}^{(1)}), \ldots, ( \sigma_{\rho}^{(7)}, \sigma_{c_{s}}^{(7)})\}$. Given the above, in order to evaluate the performance of the network, we consider as the accepted (``successful") predictions those which agree with the ``true" ones (features) within $2\sigma$ deviations
\begin{align}
& \bar{\rho}^{(J)} - 2\sigma_{\rho}^{(J)}  \leq \rho^{(J)}_{\text{true}}  \leq \bar{\rho}^{(J)} + 2\sigma_{\rho}^{(J)}, \nonumber \\ 
& \bar{c_s}^{(J)} - 2\sigma_{c_s}^{(J)}  \leq c_{s \, \text{true}}^{(J)}  \leq \bar{c_s}^{(J)} + 2\sigma_{c_s}^{(J)}
\end{align} 
An example result of this process for a given data input is shown in Fig.~\ref{fig:cs-prediction} and \ref{fig:mass-cs-histograms} for a specific choice of input. For each data point we define as the accuracy $(\text{acc}_i )\equiv N_{\text{accepted}}/7$, and the total accuracy as 
$
\text{acc} \equiv \sum_i^{N} (\text{acc})_i/N,
$
with $N$ the number of data points. The choice of the $2\sigma$ range for accepting/rejecting points is clearly a choice that we make as a reasonable credibility interval. 

In view of the above evaluation pipeline, we proceeded through trial and error, and we trained our model down to a sufficiently low, optimal value for the loss function, paying particular attention to the avoidance of overfitting. We trained and tested our model on different datasets, and found that the best test accuracy was achieved with the training dataset which inputs are $30$ triplets of masses, radii and tidal numbers. Computing the accuracy of our model on our test data according to the above procedure, we found an accuracy of 
\begin{equation}
\text{mean accuracy} \pm \sigma \simeq \left( 79 \pm 16 \right) \%. 
\end{equation}
The spread of $16 \%$ around the mean accuracy is somewhat large, implying that our model performs better on some EOS than others. It is to be emphasised that the features of our training/testing dataset, i.e the $7$ values of the sound speed, were drawn randomly from a normal distribution, as explained in Sec. \ref{Sec:dataset}. Our probabilistic network, therefore, did not have sufficient information to reconstruct the posterior functions more accurately and in turn, achieve a better accuracy. We also note that the accuracy of the prediction decreases with decreasing the number of input observations in the network, i.e the number of stellar mass, stellar radius and tidal Love number triplets. This we have verified trying with a dataset where each input to the network consisted of $15$ and $30$ pairs respectively. It is to be further noted that, the accuracy of the network's predictions on test data may increase with decreasing training accuracy. This is because the error bars of output statistical samples will be larger at lower training accuracy, a situation which is clearly undesirable. We found that an optimal choice for the training accuracy is around $70\%$, which is a fair compromise between the previous issue, and the also undesirable problem of overfitting. At this point, we should emphasize that the quoted training accuracy is according to the way that {\it TensorFlow} translates the value of the loss function to an accuracy value. In our Bayesian setting, however, the accuracy cannot be derived by simply comparing the network's single prediction with the true values of the features, but through a statistical analysis of the respective output samples.

We notice here that our Bayesian network may output predictions for the features (value of the vacuum energy shift and pairs of mass density and sound speed) which are not be within the (normalisation) range $[0,1]$ used in the training labels. This is to be expected, since the network's output is unconstrained. Within our statistical approach, what matters is that the expected values lie within a given uncertainty ($\sigma$) level around some central value.  Nevertheless, as an alternative approach, we tried to impose the sigmoid activation function in the last layer of our network. We found that the probabilistic network faced convergence issues that prevented reaching any decent value for the loss (or accuracy), so we did not consider this choice any further. 

The predictions of the Bayesian deep network for the EOS, imprinted through the scaling of the speed of sound with the mass density, show qualitatively a similar behaviour with the predictions of the deterministic network presented earlier. This is re-assuring for the consistency of our approach. The Bayesian network, however, allows for a more direct capture of the epistemic uncertainties in the input data, and a statistical analysis of the results through the production of different output realisations of the same input data. Our statistical pipeline explained earlier, provides the means to perform such an analysis.

We now discuss the predictions of the Bayesian network for the vacuum energy value in the core of the star, which provides a test of a QCD phase transition. As explained earlier, this offers a great opportunity to use neutron stars as a potential probe of the cosmological constant problem \cite{Ventagli:2024cho}. We proceed in a similar fashion as for the prediction of the sound speed. Given an input of mass, radius and tidal number, the network predicts a value of the vacuum energy. For a given input, we produce a statistical sample through repeated evaluations of the network. For each of the resulting samples we compute a mean value and the respective standard deviation, and we accept it as successful if the latter value lies within the requires $\sigma$-level. We find that we reach about $\sim 70 \%$ testing accuracy if we allow successful samples to lie within approximately $2 \sigma$ range. This is somewhat higher than the required $\sigma$-level for the prediction of the speed of sound above, but still satisfactory for our purposes. As for the deterministic model, the network is able to predict better lower vacuum energies compared to higher ones. An example of this can be seen in the histograms of Fig. \ref{fig:Lambda-histograms}, which correspond to statistical predictions of the network's output for two extreme values of the ``true" vacuum energy. The lower performance of the network for higher values of the vacuum energy is certainly related to the fact that the training set is biased towards lower values of vacuum energy, as these values tend to yield a larger set of physically acceptable solutions of the gravity equations.

\section{Employing real observational data} \label{Sec:RealData}

\begin{figure}[!tbp]
\begin{center}
    {\includegraphics[width=0.5\linewidth]{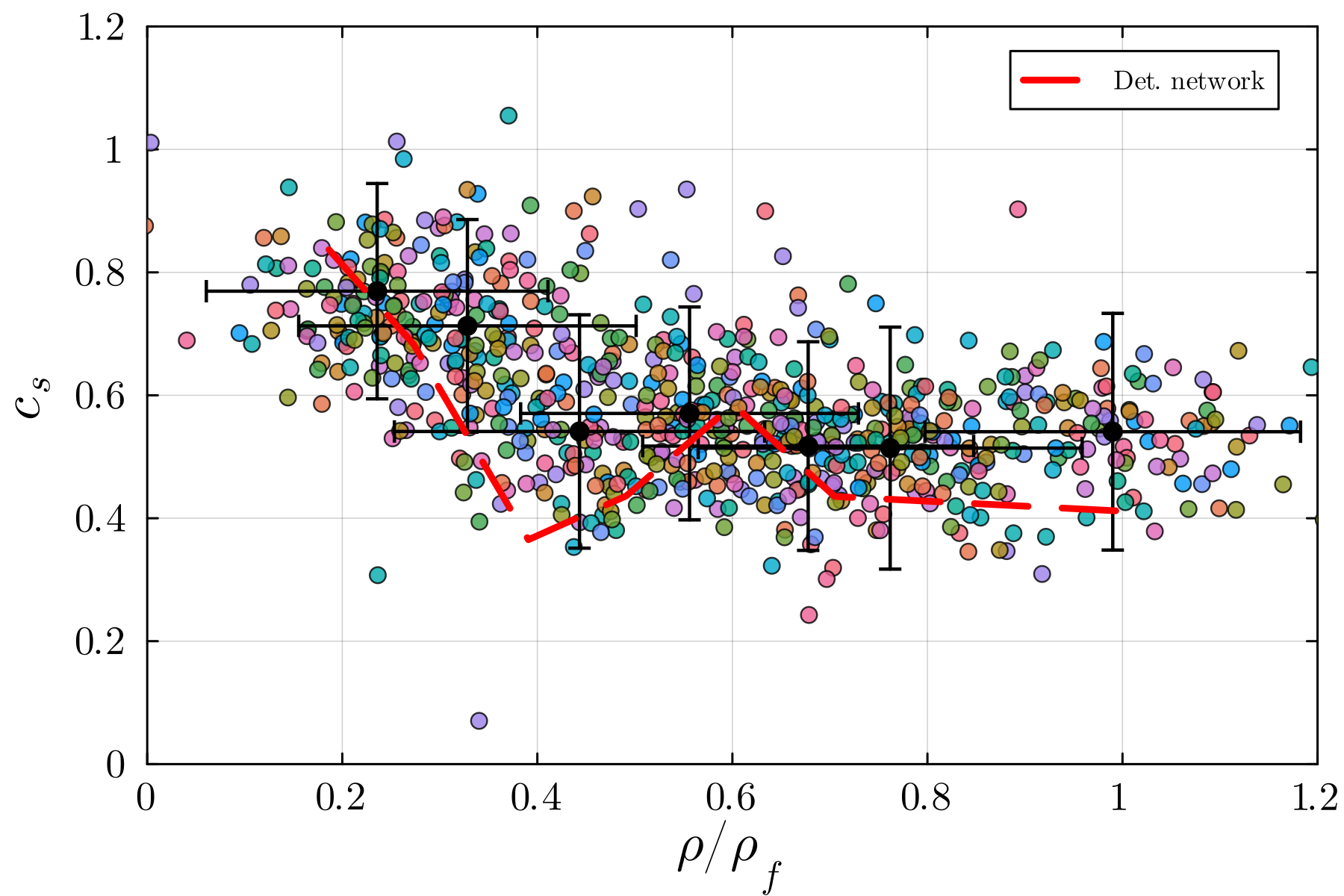}}\hfill
    {\includegraphics[width=0.5\linewidth]{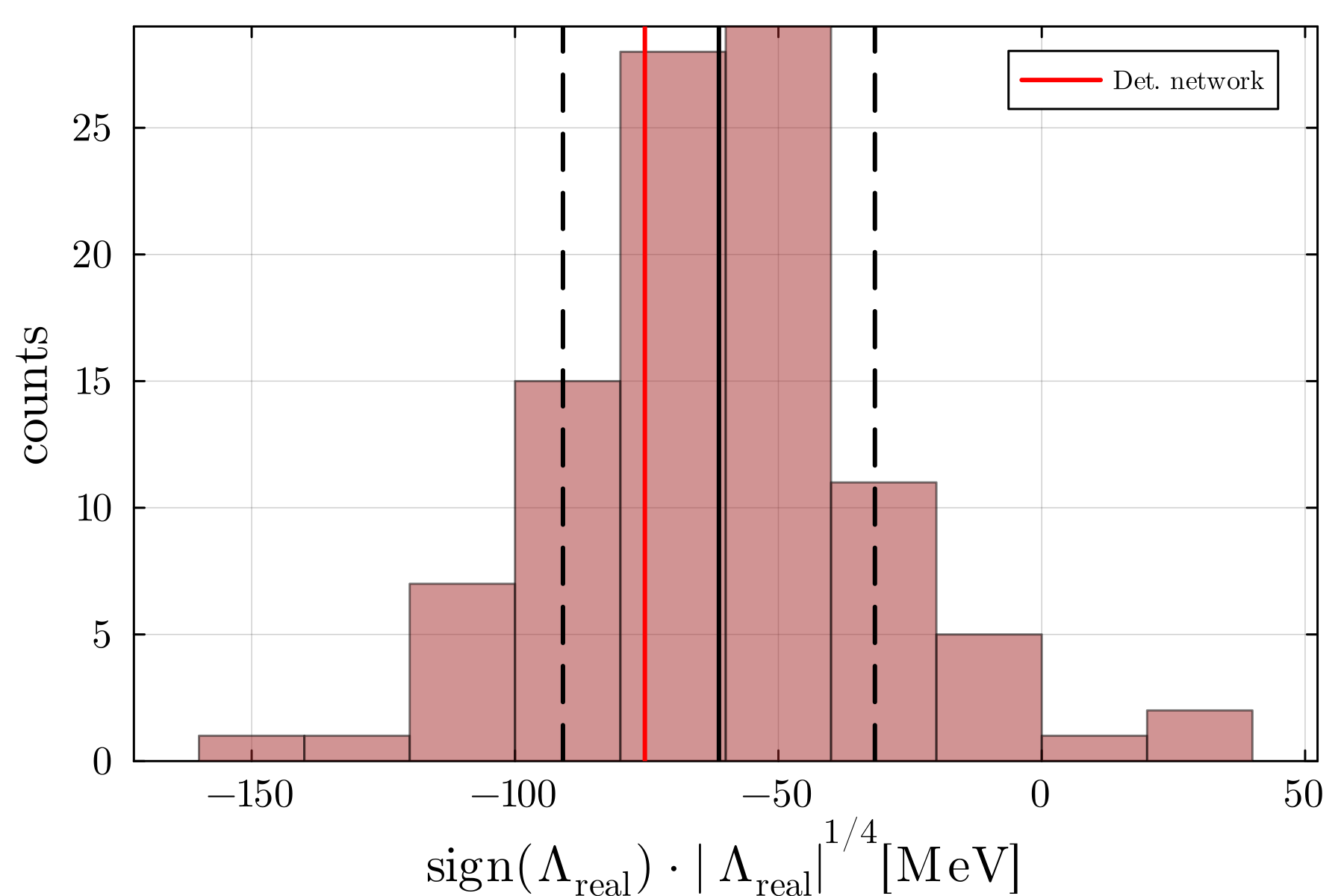}}
    \caption{Left panel: statistical sample of the predictions as outputted from the Bayesian deep network for $n_\text{samples}=100$ realisations of the output (coloured points). The black dots and the associated error bars are computed as the average and $2\sigma$ error bars respectively, given the statistical sample for each column of the output’s array. We also show the speed-of-sound profile as outputted from the deterministic model (red dashed line). We can see that the results from the deterministic network are in agreement within $2\,\sigma$ with those from the probabilistic one. We note that both the Bayesian and the deterministic deep networks predict a maximum value of speed-of-sound $c_s\simeq 0.8$ at $\rho/\rho_f\simeq0.2$. Such large value of speed-of-sound seems to be less favourable from previous results in the literature, see e.g. Ref.~\cite{Altiparmak:2022bke} where they compute the probability density function of $c_s^2$. We found that our networks' outputs consistently show a bias towards large maximum values of $c_s$ at $\rho/\rho_f\simeq0.2$ during the training process for all cases considered. Of course, the large value at $\rho/\rho_f\simeq0.2$ refers to the computed mean of the sound speed, and the corresponding error bar needs be accounted for too. This large mean value appears nevertheless to be somewhat intriguing and deserves a further investigation in the future. Right panel: predictions for the vacuum energy associated with the phase transition in the stellar core. The histograms are based on the predictions of the Bayesian deep network, for $n_\text{samples}=100$ realisations of the output. We also show the mean value with a $1\,\sigma$ interval predicted from the probabilistic network, i.e. $\text{sign}(\Lambda)\cdot|\Lambda|^{1/4} = -61\pm29\, \text{MeV}$, represented by a black solid and dashed line respectively. We include the result from the deterministic network, i.e. $\Lambda=-(75\, \text{MeV})^4$, represented by a red line. The predictions from the deterministic network are in agreement with those from the probabilistic one. Notice that the approximate $2 \sigma$ value of $\Lambda$ is practically compatible with $\Lambda = 0$. However, given the simplifying assumptions of our network and the small number of real observations, these results need to be treated with extreme care. } \label{fig:Real}
    \end{center}
\end{figure}

We now apply our networks to analyze real observational data. Since we are currently lacking varied and precise observations on the tidal deformability\footnote{The current constraint on the tidal deformability from GW170817 for a neutron star with $1.4\,M_\odot$ are $\lambda(1.4\,M_\odot)\le 800$, assuming a low-spin prior~\cite{LIGOScientific:2018hze}.}, we restrict to the analysis of masses and radii only. We use measures from X-ray observations, and we consider two different sources according to their catalogued names:

\begin{itemize}
    \item 6 thermonuclear bursters: 4U 1608-52, 4U 1724-207, KS 1731-260, EXO 1745-248, SAX J1748.9-2021 and 4U 1820-30~\cite{Ozel:2015fia,Ozel:2016oaf}.
    \item 2 pulsars from NICER: PSR J0030+0451~\cite{Miller:2019cac} and PSR J0740+6620~\cite{Miller:2021qha}.
\end{itemize}
From these observations we randomly sample 15 $(M,R)$ points and we appropriately normalized them by $M_\text{norm}=3\,M_\odot$ and $R_\text{norm}=20\,\text{km}$, as described in Section~\ref{Sec:dataset}. We used these data as input for both the deterministic regression model and the Bayesian network, discussed in Sections~\ref{Sec:deterministic} and~\ref{Sec:Bayesian}. We show the results in Fig.~\ref{fig:Real}. On the left panel, we show a statistical sample of the predictions as outputted from the Bayesian deep network for $n_\text{samples}=100$ realisations of the output (coloured points). The black dots and the associated error bars are computed as the average and $2\sigma$ error bars respectively, given the statistical sample for each column of the output’s array. We also show the speed-of-sound profile as outputted from the deterministic model (red dashed line). We can see that the results from the deterministic network are in agreement within $2\,\sigma$ with those from the probabilistic one. On the right panel, we show the predictions for the vacuum energy associated with the phase transition in the stellar core. The histograms are based on the predictions of the Bayesian deep network, for $n_\text{samples}=100$ realisations of the output. We also show the mean value with a $1\sigma$ interval predicted from the probabilistic network, i.e. $\text{sign}(\Lambda)\cdot|\Lambda|^{1/4} = -61\pm29\, \text{MeV}$, represented by a black solid and dashed line respectively. We include the result from the deterministic network, i.e. $\Lambda=-(75\, \text{MeV})^4$, represented by a red line. We can see that the predictions from the deterministic network are in agreement with those from the probabilistic one. However, we refrain from making any claims on the value of $\Lambda$, since as we saw during the training and testing of our models, the networks cannot confidently constrain this value. Improving both the training dataset and the network can potentially shed some light on possible constraints on the value of $\Lambda$. We leave this for future work. Finally, we also note that the approximate $2\,\sigma$ value of $\Lambda$ is practically compatible with $\Lambda = 0$.

\section{Conclusions} \label{Sec:Conclusions}

We have constructed a pipeline for inferring the equation of state (EOS) of neutron stars with deep neural networks, based on conventional (deterministic) and Bayesian (probabilistic) architectures. Our modelling of the EOS in the stellar interior is composed of three regions. At low densities, which corresponds to the most well-understood region, we employed a tabulated EOS (AP4 or SLy), while at high densities we used an agnostic speed-of-sound parametrization of the EOS according to Ref.~\cite{Tews:2018iwm}. Most importantly, we also allowed for a vacuum energy phase transition at the stellar core, as prescribed in Ref.~\cite{Ventagli:2024cho}, a key novelty of our work. The latter opens up the path to use neutron star observations to probe the nature of the cosmological constant.
Indeed, this shift in energy can be interpreted as a new effective cosmological constant term $\Lambda$ contributing to the pressure, energy density and mass density of the new exotic phase in the stellar core.
Measurements of the imprints of such transition on the stellar properties have the potential to provide a first test of the gravitational properties of vacuum energy independent from the acceleration of the Universe.
Our training and validation datasets were based on a simulated set of $1491$ EOSs, from which we subsequently constructed sets of $N=15,\,20,\,30$ mock observations of total stellar mass, stellar radius and tidal deformability respectively. For each set of mock data, we injected Gaussian noise to simulate observational uncertainties, and repeated this procedure for $n=100,\,200,\, 300$ times.

For the conventional (deterministic) deep network, we developed two separate pipelines, namely, a classification model to distinguish the baseline EOS at low density (AP4 or SLy), and a regression model to predict the values of the vacuum energy shift and speed-of-sound profile in the high density region. Overall, performance of all our networks on testing data increased with increasing the number of data at each input. The inference of the low-density EOS, done through the classification model, reached a testing accuracy of $\sim 87.0\%$. However, the high-density part of the EOS, inferred through the regression model, appears not to be in agreement with the expected behavior, as shown in Fig.~\ref{fig:csProfile}. This can also be seen from the analysis of the predicted mean speed-of-sound for all possible EOS' configurations shown in Fig.~\ref{fig:csMean}. The deep network tends to smooth out the speed of sound profile, missing to capture the oscillatory pattern. This should not come as a surprise, since the number of features we provide during training (at fixed input) are not only limited (7 pairs of points), but they are also sampled randomly according to our explained parametrization. It is, therefore, hard for the network to catch the finer features of the sound speed profile. When considering predictions on the vacuum energy shift, the network's output provided a better agreement with the expected/predicted values, as shown in Fig.~\ref{fig:LambdaMean}. Almost all predicted values lie within a $2\,\sigma$ interval, with the exception of very large positive values of $\Lambda$. The reason why our network can better identify a large negative vacuum energy, rather than a (large) positive one, lies on the different properties of these two cases, which can be considered as two distinct families of EOSs~\cite{Ventagli:2024cho}: configurations which admit a wide range of values for $\Lambda$, and those which are allowed only if a large negative jump in vacuum energy is introduced. This leads to different stellar properties. In particular, the second family displays peculiar $M$-$R$ relations, and the network can better distinguish them between all possible configurations. Note also that our training dataset is biased towards lower values of the vacuum energy jump, since these choices lead to a larger set of physically acceptable solutions of the gravity equations.

For the Bayesian deep network, we focused on the regression problem of predicting the sound speed profile and the vacuum energy shift. Our network architecture employed variational layers, where each weight and bias was drawn from a learnable posterior distribution. For a reasonably simple network, our best accuracy on testing data  for the sound speed profile was $(79\pm16)\%$, assuming that the network's predictions agree with the expected values within $2\,\sigma$ deviations. The large spread around the mean accuracy implies that our model performs better on some EOSs than others. Crucially, the predictions of the probabilistic network show qualitatively a similar behaviour with the predictions of the deterministic model, i.e. the speed of sound profile tends to be smoothed out missing large oscillations on the value of sound speed. This can be seen in Fig.~\ref{fig:cs-prediction}. The Bayesian network, however, allows for a more direct capture of the epistemic uncertainties in the input data, and a statistical analysis of the results through the production of different output realisations of the same input data. When focusing on the vacuum energy shift, the network testing accuracy reaches about $\sim 70 \%$, assuming that predictions agree with the expected value within $2\,\sigma$ deviations. 

Finally, we applied our network to real mass and radius data from X-ray observations. We used these data as input to both our deterministic and probabilistic network. The results are shown in Fig~\ref{fig:Real}, where we show both the output for the speed-of-sound profile and that for the vacuum energy shift. The results from the deterministic network and those from the probabilistic one are in agreement. From the Bayesian network, we find that the vacuum energy jump's mean value with a $1\sigma$ interval is given by $\text{sign}(\Lambda)\cdot|\Lambda|^{1/4} = -61\pm29\, \text{MeV}$. Given the simplifying assumptions of our network and the small number of real observations, this result needs to be treated with extreme care. We also notice that the approximate $2\,\sigma$ value of $\Lambda$ is practically compatible with $\Lambda = 0$."

There are various interesting future directions for our work. First, one could allow the observational data to have a lower or varying uncertainties. We expect that such step would allow for a higher accuracy of our network. It would also be interesting to perform a similar analysis employing the machinery of normalising flows to construct the posterior of our data, thus providing a conceptually different analysis of the pipeline presented here. Our pipeline was based on a particular parametrisation for the EOS, and it would be essential to test the performance of our network on different EOS parametrisations for the high-density regime of the star. We also intend to improve the study at low-densities by either allowing for a wider range of low-density EOSs from those commonly used in the literature, or by extending the analysis of the regression model to take into account this region as well. This would allow for a more appropriate treatment of the uncertainties that are still present at low-densities. Furthermore, to reduce the bias in our training data toward lower values of the vacuum energy, it would be crucial to allow for more configurations with positive values of vacuum energy shift. Finally, a promising direction would be to complement our datasets with gravitational waveforms, this way extending our analysis to the case of neutron-star mergers. We leave these aspects for future work.

\acknowledgments
We thank the referee for their constructive feedback.
We thank the EuCAIFCon for useful discussion on this work during the conference. G.V. acknowledges support by the Czech Academy of Sciences under the grant number LQ100102101, and I.D.S by the Czech Grant Agency (GA\^CR) under the grant number 21-16583M. 
\bibliographystyle{JHEP}
\bibliography{Ref.bib}

\end{document}